\documentclass[11pt,a4paper]{article}
\usepackage{jheppub}
\usepackage[utf8]{inputenc}
\usepackage{physics}
\usepackage{tikz-feynman}
\usepackage{caption}
\usepackage{xcolor}
\usepackage{comment}
\usepackage{multirow}
\usepackage{graphics}
\usepackage{float}
\usepackage{cancel}
\usepackage{soul}
\usepackage{cancel}
\usepackage{array}

\tikzfeynmanset{compat=1.1.0}
\newcolumntype{L}[1]{>{\raggedright\let\newline\\\arraybackslash\hspace{0pt}}m{#1}}
\newcolumntype{C}[1]{>{\centering\let\newline\\\arraybackslash\hspace{0pt}}m{#1}}
\newcolumntype{R}[1]{>{\raggedleft\let\newline\\\arraybackslash\hspace{0pt}}m{#1}}

\def\nn{\nonumber}
\def\l{\left}
\def\r{\right}
\def\c{\color}

\title{\boldmath{Amplitude's positivity vs. subluminality: Causality and Unitarity Constraints on dimension 6 \& 8 Gluonic operators in the SMEFT}}

\author[a,1]{Diptimoy Ghosh}
\author[a,2]{Rajat Sharma}
\author[a,3]{Farman Ullah}

\affiliation[a]{Department of Physics\\Indian Institute of Science Education and Research Pune, India}

\emailAdd{diptimoy.ghosh@iiserpune.ac.in}
\emailAdd{rajat.sharma@students.iiserpune.ac.in}
\emailAdd{farman.ullah@students.iiserpune.ac.in}

\abstract{We derive the causality and unitarity constraints on dimension 6 and dimension 8 Gluon field strength operators in the Standard Model Effective Field Theory (SMEFT). In the first part of the paper, we use the `amplitude analysis' i.e. dispersion relation for $2\rightarrow2$ scattering in the forward limit, to put bounds on the Wilson coefficients. We show that the dimension 6 operators can exist only in the presence of certain dimension 8 operators. It is interesting that the square of the dimension 6 Wilson coefficients can be constrained in this case even at the tree level. 
In the second part of this work, we successfully rederive all these bounds using the classical causality argument that demands that the speed of fluctuations about any non-trivial background should not exceed the speed of light. We also point out some subtleties in the superluminality analysis regarding whether the low-frequency phase velocity can always be used as the relevant quantity for Causality violation: as an example, we show that, due to these subtleties, if a small pion mass is added in the chiral Lagrangian, it is unclear if any strict positivity bound can be derived on the dimension 8 Wilson coefficient. Finally, we mention an interesting non-relativistic example where the subluminality requirement produces a stronger bound than the `amplitude analysis'.}

\begin{document}

\maketitle
\section{Introduction}\label{intro}

It is well known that dynamics at very high energy scales or short distances are irrelevant to describe low energy or long distance phsyics, i.e. very different energy scales are `decoupled' from each other. For example, we don't need to know the fine details of nuclei to understand the properties of electronic energy levels in atoms. It's mainly on this idea that the framework of Effective Field Theory (EFT) is built (see \cite{burgess,Skiba:2010xn} for a review), in which one constructs the Lagrangian for the low energy (IR) theory in terms of some physical cut-off energy scale ($\Lambda$). The Lagrangian for the EFT, a priori, must contain all the possible operators consistent with the symmetries of the theory e.g., Lorentz invariance and gauge invariance, and  the coefficients of these operators can have arbitrary values. However, it has been shown in recent years that sacred principles like relativistic Causality and Unitarity do impose non-trivial constraints on these coefficients and carve out the allowed parameter space \cite{causality,DSD2,derhamcausalEFT,snowmass}. This is interesting also phenomenologically since it leads to enhanced statistical power for experiments sensitive to these operators because one can incorporate IR consistency bounds into the prior probability distribution.

One of the first attempts in this direction due to \cite{PhysRevD.31.3027,causality,posscalar,posgravity} (which we refer to as the `amplitude analysis') exploits well-established fundamental principles like micro-causality (leading to analyticity \cite{analyticity_Bros,analyticity_Lehmann}) and unitarity of the S-matrix to constrain the EFT parameter space. This involves using dispersion relations for $2\rightarrow2$ scattering amplitudes in the forward limit. The `amplitude analysis' has successfully given linear positivity bounds on dimension 8 operators in a variety of theories but hasn't had much success with dimension 6 operators\footnote{See recent developments made to constrain dim\hspace{1.5pt}6 operators using S-matrix Bootstrap methods \cite{dim6_Elias_miro}.} containing 4 fields; although, one can derive certain sum rules \cite{0907.5413,1202.1532,1807.06940,2008.07551,2010.04723,2206.13524,dim6_amartya}. The reason for this lack of success is  the fact that for such an operator, the scattering amplitude grows as the Mandelstam variable $s$ at the tree level. However, in this article, we show that \textit{it's possible to constrain the square of the coefficient of dimension 6 operators containing  \textbf{3} gluon fields  w.r.t those of dim 8 operators}. Such an operator appears in the SMEFT. Similar positivity bounds for the electroweak gauge bosons were 
obtained in \cite{constraintvectorboson}. A lot of effort has been focused on constraining the parameter space of SMEFT using various methods (see, for example, \cite{Zhou_positivity_multifields,constraintvectorboson,Flavor_constraints,minimal_flovor_violation,Trott_EFT,consistentSMEFT} and the references therein), since it is and will be the main aim of present and future particle physics experiments to  measure these coefficients. Also, verifying whether the experimentally measured coefficients satisfy our theoretical constraints allows us to test fundamental properties of the UV theory such as locality and Lorentz invariance up to very high energies through experimental signatures at accessible scales \cite{consistentSMEFT}.

The other method often employed to put constraints on the Wilson coefficients is based on the classical causality argument \cite{causality}. One demands that the propagation of perturbations over any non-trivial background should respect causality i.e. the speed of signal propagation should not be superluminal \cite{causality,snowmass,derhamcausalEFT,Goon}.  It is well known that if the wavefront velocity i.e. infinite frequency limit of the phase velocity is (sub)luminal then causality is preserved. Naively, one might conclude that it is not possible to put constraints on the EFT coefficients since the EFT is valid, by definition, in the low-frequency regime i.e. $\omega/\Lambda\ll1$. However, one can instead consider signal velocity (for a precise definition, see \cite{BrillouinLeon1960Wpag}) which, for non-dispersive mediums (the case of our interest), is equal to the group/phase velocity. Thus, the low frequency group/phase velocity can be directly associated with causality obviating the need to take the high-frequency limit. One can also use analyticity in the form of the Kramers-Kronig relation \cite{Jackson:1998nia}, which, for dissipative backgrounds, demands that the phase velocity cannot decrease with increasing frequency \cite{Jackson:1998nia,snowmass}. Therefore, the superluminal phase velocity in the EFT can be associated with causality violation. However, in general, it is not very clear how one can determine the dispersive properties of the background medium, which is essential in the usefulness of the Kramers-Kronig relation. We will discuss this in some more detail in section~\ref{superluminality}.

It is not always necessary that \textit{small} superluminal low energy speed violates causality as the observations detecting causality violation may turn out to be unmeasurable within the valid regime of EFT \cite{Goon,Drummond}. Therefore for generic EFTs, particularly gravitational ones, scattering phase shift or time delay is perhaps a better probe to detect causality violations \cite{CEMZ,timedelay}. However, for homogeneous backgrounds (as considered in this paper) signals can be allowed to propagate over large distances, and in that case, even the small superluminality can be detected within the EFT regime. Therefore, using this method one can try to rederive or even hope to improve the bounds on EFT coefficients obtained by the `amplitude analysis'.

We would like to stress that, a priori, it is unclear if the two methods always provide the same constraints (or equivalently, whether using any one of them is enough to maximally constrain the space of EFTs), as naively, they don't seem to be related at all. One is purely based on the classical causality of wave propagation and another on scattering amplitudes which relies on Unitarity, and Froissart bound in addition to micro-causality (the two methods could be somewhat related since the classical causality analysis secretly might also depend on analyticity in the form of Kramers-Kronig relation \cite{snowmass} and unitarity (for dissipative mediums) however, the connection is unclear, as we discuss in section~\ref{superluminality}).

For the classical causality/superluminality analysis, we also point out some subtleties that arise when mass-like terms exist in the dispersion relation. We show, in the particular case of the chiral Lagrangian, that this may lead to deviation from strict positivity. However, in the case of gluonic operators, we demonstrate the mass-like terms can be removed by choosing particular configurations of non-trivial background and polarization of perturbation. This helps us derive constraints on the Wilson coefficients of gluonic operators by demanding subluminal phase velocity as our measure for causality.  We show that \textit{the superluminality analysis for dim\hspace{1.5pt}6 and 8 gluonic operators, in a nontrivial way, reproduces all bounds that we obtain from the `amplitude analysis'}. This is the novel and main result of our work. Finally, we mention a non-relativistic example following \cite{Baumann_2016}, where superluminality gives stronger bounds than the amplitude analysis.
From the examples given in our work, one can gather that one should use both analyses whenever possible in order to get the maximum amount of information on an IR effective theory\footnote{For fermions, it is unclear how one can implement the superluminality analysis. However, the `amplitude analysis' can still be carried out \cite{fermions_ampl}.}. 

The rest of the paper is organized as follows: In section~\ref{notation}, we define all the notations and conventions we have used throughout the paper to avoid any confusion. In section~\ref{sec3}, we derive positivity constraints on dim\hspace{1.5pt}6 and dim\hspace{1.5pt}8 gluonic operators in SMEFT using the `amplitude analysis' with an overview of the method first. In section~\ref{superluminality}, we first discuss a few subtleties in the superluminality analysis, followed by a demonstration that all the bounds can also be reproduced by the superluminality analysis. The summary of our results is presented in section~\ref{summary}.

\section{Notation and Conventions}\label{notation}
We use $\eta_{\mu\nu}=(+,-,-,-) $ metric and work in Lorentz gauge.

\vspace{1.8mm}
\noindent Greek indices $\mu, \nu$, etc. run over four space-time coordinate labels 0,1,2,3 with $x^0$ being the time coordinate.

\vspace{1.8mm}
\noindent Repeated indices are summed over unless otherwise specified.

\vspace{1.8mm}
\noindent The complex conjugate and hermitian adjoint of a vector or a matrix A are denoted by $A^*$ and $A^{\dag}$.

\vspace{1.8mm}
\noindent The gluon field strength tensor is defined as \quad
$\displaystyle{G^{a}_{\mu\nu}=\partial_\mu A^{a}_\nu-\partial_\nu A_\mu^a+g_{s}  f^{abc}A^b_\mu A_\nu^c}$ \quad and its dual tensor as \quad
$\displaystyle{\widetilde{G}^{a}_{\mu\nu}=\frac{1}{2}\varepsilon_{\mu\nu\rho\sigma}G^{a\;\rho\sigma}\;\; (\varepsilon_{0123}=+1)}$ \quad
where
$A_\mu =t_{a}A_\mu^{a}$ expresses the gluon field; $t_a$ are the generators of SU(3); $a$, $b$, $c$ = 1, 2, ..., 8 are color indices; $g_s$ is the coupling constant of the strong force; $f^{abc}$ and $d^{abc}$ are the total anti-symmetric and symmetric structure constants of SU(3) respectively, defined as
$$\l[t^a,t^b\r]=if^{abc}t^c\;;\;\quad \l\{t^a, t^b\r\}=\frac{1}{3} \delta^{a b}+d^{a b c} t^c$$

\vspace{1.8mm}
\noindent For convenience we define \quad $\displaystyle{F^a_{\mu\nu}=\partial_\mu A^{a}_\nu-\partial_\nu A_\mu^a}$.

\vspace{1.8mm}
\noindent In scattering amplitudes, we consider $p_1$ and $p_2$ as incoming momenta and $p_3$ and $p_4$ as outgoing, and define Mandelstam invariants:

\hfil $s=(p_1+p_2)^2 , \quad t=(p_1-p_3)^2 , \quad u=(p_1-p_4)^2$ \\
satisfying $s+t+u=m_1^2+m_2^2+m_3^2+m_4^2$ where $m_i$ are masses of the particles.

\section{Positivity constraints from unitarity and analyticity}\label{sec3}
In this section, we derive constraints on dimension 6 and dimension 8 Gluonic operators of the SMEFT using dispersion relations for $2\rightarrow2$ scattering amplitude. Let us first review
how the `amplitude analysis' works to put constraints on the Wilson coefficients of a general EFT.

\subsection{Overview}\label{s3}
In this section, we give a brief overview, following the discussion in the seminal paper \cite{causality}, of how the dispersion relations along with well-established principles like micro-causality (leading to analyticity of S-matrix), unitarity, and locality, can help in constraining the low-energy EFT parameters which from the IR side can have arbitrary values, to begin with. We slightly modify the discussion in accordance with our work but the main underlying concepts are the same.

\begin{figure}
    \vspace{-1cm}
    \centering
    \includegraphics[scale=0.6]{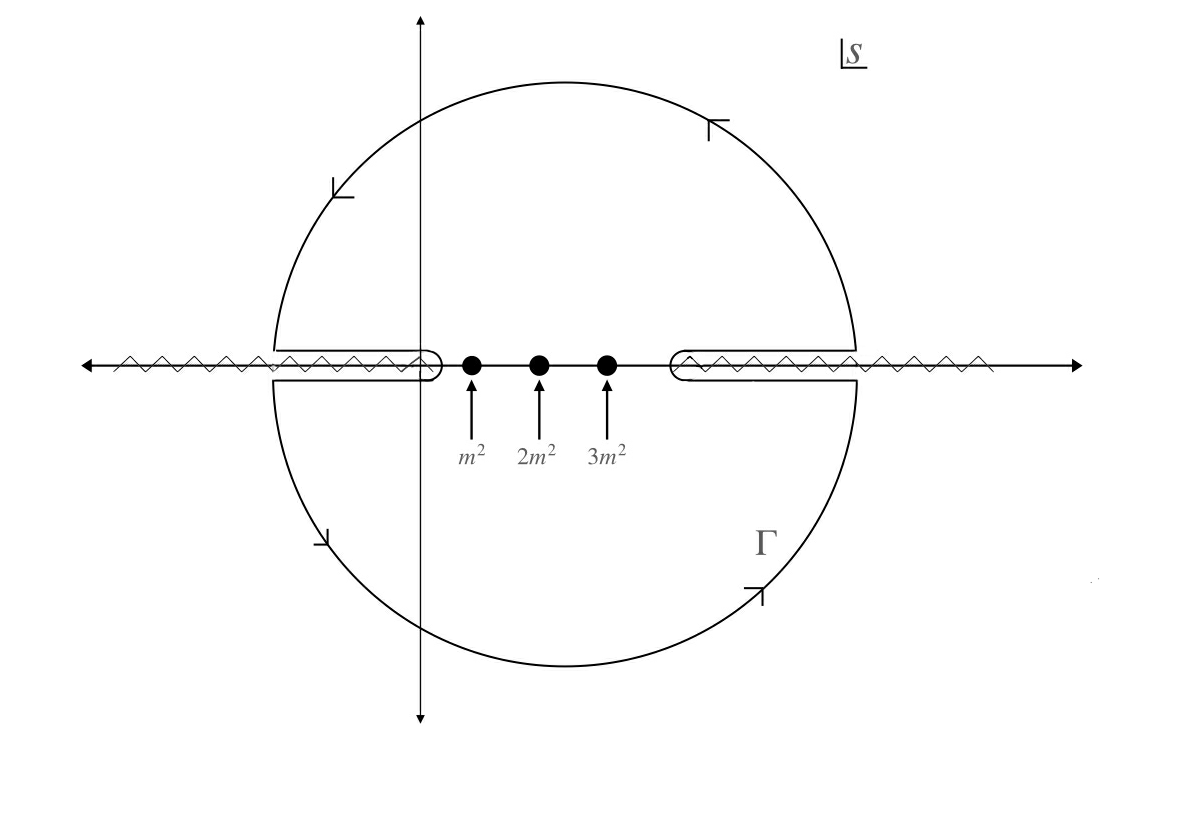}
    \caption{Analytic structure of $\mathcal{A}(s)$ in complex s plane. The contour is symmetric about $s=2m^2$ and go over branch cuts from $-\infty$ to 0 and $4m^2$ to $\infty$.}
    \label{contour}
\end{figure}

\vspace{2mm}
Let us consider $s\leftrightarrow u$ symmetric $2\rightarrow2$ scattering amplitude, $\mathcal{M}(s,t)$ for a process in which exchanged particles or particles inside a loop are of the same mass $m$ (or massless as would be the case in our work), in the forward limit i.e. initial and final states are exactly same.  We define $\displaystyle{\mathcal{A}(s)=\mathcal{M}(s,t)|_{t\rightarrow0}}$ (forward limit) and take integral around contour as shown in fig.~\ref{contour}
\begin{equation}\label{3.1}
    \mathcal{I}=\oint_C\frac{ds}{2\pi i} \frac{\mathcal{A}(s)}{(s-2m^2)^3}
\end{equation}
where $m$ is the mass of the exchanged particles (or regularized mass for massless exchanged particles), and $\Lambda$ is the energy cut-off scale of EFT in consideration.
 It is possible that the theory considered allows for the loops leading to branch cuts starting from $-\infty$ to $0$ and $4m^2$ to $+\infty$. That is why we probe the $s\rightarrow 2m^2$ limit instead of $s\rightarrow0$ (which would be okay in case the branch cut doesn't go through or extend up to 0). 
 \\
 Now we evaluate the integral (\ref{3.1}); since $\lim_{|s|\to \infty} \mathcal{A}(s)/|s|^2 \to 0$ at infinity due to the Froissart bound (more precisely, $\mathcal{A}(s)<s\;ln^2s$) \cite{froissart,jim-martin}, the integral over the arc at infinity vanishes and we are left just with the integral of discontinuity of $\mathcal{A}(s)$ across the branch cuts, 
 $$\mathcal{I}=\frac{1}{2\pi i}\int_{cuts}ds\frac{Disc\mathcal{A}(s)}{(s-2m^2)^3}$$
 But the integral can also be evaluated in terms of the residues at the  poles: at $s=2m^2$ and $s=m^2,\;3m^2$ (due to $s$-channel and $u$-channel in exchange diagrams). Thus, we get
 \begin{equation}
     \frac{1}{2}\mathcal{A}''(s=2m^2)+\sum_{s^*=m^2,3m^2}\frac{res\mathcal{A}(s=s^*)}{(s^*-2m^2)^3} = \frac{2}{2\pi i}\int_{s>4m^2}ds\frac{Disc\mathcal{A}(s)}{(s-2m^2)^3}
 \end{equation}
In the above equation, we have mapped the integral over the negative branch cut to the positive one using $s\leftrightarrow u$ crossing symmetry.
Now, $Disc\mathcal{A}(s)=2i Im\mathcal{A}(s)$ and from optical theorem (for which initial and final states are required to be identical), we have $Im\mathcal{A}(s)=\sqrt{s(s-4m^2)}\sigma(s)$ where $\sigma(s)$ is the total cross-section of the scattering,
 \begin{align}\label{3.3}
 \frac{1}{2}\mathcal{A}''(s=2m^2)+\sum_{s^*=m^2,3m^2}\frac{res(\mathcal{A}(s=s^*))}{(s^*-2m^2)^3}=\frac{2}{\pi}\int_{s>4m^2}ds\frac{\sqrt{s(s-4m^2)}\sigma(s)}{(s-2m^2)^3}
 \end{align}
 For further analysis, we'll take the $s\leftrightarrow u$ symmetric $\mathcal{A}(s)$ to be of a particular form, which we'll be encountering in further sections.

 We concern ourselves with operators only up to dimension 8 in the low-energy EFT and also assume that it contains only 6 and 8 dimension operators in addition to 4-dimensional terms.
 If the theory allows taking $t\rightarrow0$ i.e. forward limit without causing any divergence problem (which will be the case here) then we can write the $s\leftrightarrow u$ symmetric forward scattering amplitude at tree level as 
\begin{align}\label{3.4}
    \mathcal{A}(s)=\lambda+b\frac{m^2}{\Lambda^2}&+\frac{1}{\Lambda^4}(c_1s^2+c_1u^2+c_3m^4)\\\nonumber
    &+\frac{1}{\Lambda^4}\left(c_4\frac{s^3}{s-m^2}+c_5\frac{s^2m^2}{s-m^2}+c_6\frac{sm^4}{s-m^2}+c_7\frac{m^6}{s-m^2}\right)\\
    &+\frac{1}{\Lambda^4}\left(c_4\frac{u^3}{u-m^2}+c_5\frac{u^2m^2}{u-m^2}+c_6\frac{um^4}{u-m^2}+c_7\frac{m^6}{u-m^2}\right)\nonumber
\end{align}
Note: If we take the exchange particles to be massless, then it might not be possible to take the forward limit, even with the  regularized mass because we also need to put $m\rightarrow0$ at some stage, which is one of the main problems in performing this analysis for EFTs of gravity. Then for the massless case, either the t-channel shouldn't exist or the numerator in the t-channel exchange contribution should converge to 0 faster than t in $t\rightarrow0$ limit avoiding the t-channel pole problem which would be the case for our EFT in consideration.\\
Putting (\ref{3.4}) in (\ref{3.3}) we get,
\begin{equation}\label{positivity}
    \frac{2}{\Lambda^4}\left(c_1+c_4\right)=\frac{2}{\pi}\int_{s>4m^2}ds\frac{\sqrt{s(s-4m^2)}\sigma(s)}{(s-2m^2)^3}
\end{equation}
%
The integrand in the r.h.s of the above equation is positive as the cross-section $\sigma(s)>0$, which then makes the r.h.s manifestly positive. Therefore, the above equation shows that the coefficient of the term which goes as $s^2$ upon taking $m\rightarrow0$ limit of $\mathcal{A}(s)$ (l.h.s of the above equation) is positive. The same result can be obtained by using a different contour as shown in appendix~\ref{arc}.

%

\subsection{Relative bounds on dim\hspace{1.5pt}6 and dim\hspace{1.5pt}8 operators}\label{amp}

It is clear from the previous section that we cannot put any constraint on the contribution that grows slower than $s^2$. This is because for the integral to vanish over the arc at infinity,  we need minimum $n=3$ in $\displaystyle{\oint_c\frac{\mathcal{A}(s)}{(s-2m^2)^n}}$ but then there is no contribution to the residue from terms growing less than $s^2$, preventing us from constraining them. If we take $n=2$ and even ignore the contribution from arcs at infinity, then we get the contribution from dim 6 operators towards the residue in $\text{eq}^\text{n}$(\ref{positivity}). However, for $n=2$ the integrand on r.h.s of (\ref{positivity}) takes form $\sigma(s)/s$ which makes the integral have a non-definitive sign. Thus, the l.h.s also have a non-definitive sign preventing us from constraining the Wilson coefficients using positivity arguments.

\vspace{3mm}
\noindent This is usually the case for dimension 6 operators containing  four fields as their contribution to the tree level amplitude grows as  $s$. However, if the dim\hspace{1.5pt}6 operator also gives rise to terms containing only three fields then one indeed gets an $s^{2}$ piece through exchange diagrams. The constraint in this case, however, is on the square of the Wilson coefficient (and not on the sign). Also, dimension 8 operators containing four fields give contributions at the same order ($\displaystyle{s^2/\Lambda^4}$) via contact diagrams. Thus, we produce relative bounds on dim\hspace{1.5pt}6 and dim\hspace{1.5pt}8 operators. Similar bounds for the electroweak gauge bosons were 
obtained in \cite{constraintvectorboson}.

\begin{figure}
     \centering
     \includegraphics[scale=0.32]{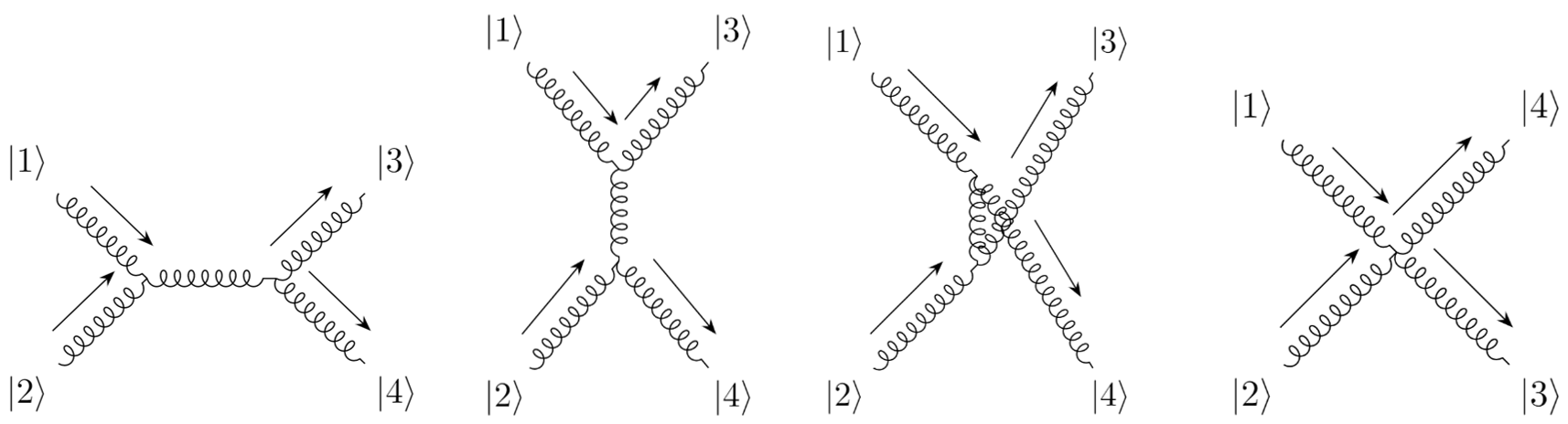}
     \caption{Exchange diagrams (first three) get contribution just from dim\hspace{0.4mm}6 operators ; contact diagram gets contribution from both dim\hspace{0.4mm}6 and dim\hspace{0.4mm}8 but the dim\hspace{0.4mm}6 contribution is of order $s/\Lambda^2$ as explained in the text.}
     \label{gluon}
 \end{figure}

\vspace{2mm}
Let us now apply the above analysis for dimensions 6 and 8 Gluonic operators in the SMEFT.
 The independent Gluonic operators for SU(3) as mentioned in \cite{dim6} and \cite{dim8,dim_8_008} are given in table~\ref{dim6_dim8_operators}.
\begin{table}[h]
\centering
\resizebox{12cm}{!}{
\begin{tabular}{||c|c||c|c||}
\hline \multicolumn{2}{||c||}{$X^{3}$} & \multicolumn{2}{c||}{$X^{4}$} \\
\hline
$Q_{G^3}^{(1)}$ &$f^{abc} G_\mu^{a\nu}G_\nu^{b\rho}G_\rho^{c\mu}$&$Q_{G^4}^{(1)}$ & $\left(G_{\mu \nu}^{a} G^{a \mu \nu}\right)\left(G_{\rho \sigma}^{b} G^{b \rho \sigma}\right)$   \\
$Q_{G^3}^{(2)}$ & $f^{abc} \widetilde{G}_\mu^{a\nu}G_\nu^{b\rho}G_\rho^{c\mu}$&$Q_{G^4}^{(2)}$ & $\left(G_{\mu \nu}^{a} \widetilde{G}^{a \mu \nu}\right)\left(G_{\rho \sigma}^{b} \widetilde{G}^{b \rho \sigma}\right)$  \\
 &  & $Q_{G^4}^{(3)}$ &  $\left(G_{\mu \nu}^{a} G^{b \mu \nu}\right)\left(G_{\rho \sigma}^{a} G^{b \rho \sigma}\right)$\\
 &  & $Q_{G^4}^{(4)}$ & $\left(G_{\mu \nu}^{a} \widetilde{G}^{b \mu \nu}\right)\left(G_{\rho \sigma}^{a} \widetilde{G}^{b \rho \sigma}\right)$ \\
 &  &$Q_{G^4}^{(5)}$ & $\left(G_{\mu \nu}^{a} G^{a \mu \nu}\right)\left(G_{\rho \sigma}^{b} \widetilde{G}^{b \rho \sigma}\right)$ \\
&  & $Q_{G^4}^{(6)}$ & $\left(G_{\mu \nu}^{a} G^{b \mu \nu}\right)\left(G_{\rho \sigma}^{a} \widetilde{G}^{b \rho \sigma}\right)$ \\
 &  & $Q_{G^4}^{(7)}$ & $d^{a b c} d^{d e c}\left(G_{\mu \nu}^{a} G^{b \mu \nu}\right)\left(G_{\rho \sigma}^{d} G^{e \rho \sigma}\right)$\\
 &  & $Q_{G^4}^{(8)}$ & $d^{a b c} d^{d e c}\left(G_{\mu \nu}^{a} \widetilde{G}^{b \mu \nu}\right)\left(G_{\rho \sigma}^{d} \widetilde{G}^{e \rho \sigma}\right)$\\
 &  & $Q_{G^4}^{(9)}$ & $d^{a b c} d^{d e c}\left(G_{\mu \nu}^{a} G^{b \mu \nu}\right)\left(G_{\rho \sigma}^{d} \widetilde{G}^{e \rho \sigma}\right)$\\
\hline
\end{tabular}
}
 \caption{Dimension 6 and 8 gluonic operators in the SMEFT}
 \label{dim6_dim8_operators}
\end{table}
 The Lagrangian at dimensions 6 and 8 takes the following form:
 $$L^{(6)}=\frac{c_6}{\Lambda^2}Q_{G^3}^{(1)}+\frac{c_6'}{\Lambda^2}Q_{G^3}^{(2)}\;;\;\;\; L^{(8)}=\frac{c_8^{(i)}}{\Lambda^4}Q_{G^4}^{(i)}$$
 where $c_6$, $c_6'$ and $c_8^{(i)}$ are dimensionless Wilson coefficients and $\Lambda$ is the UV cut-off for the EFT.
 The above-mentioned dimension 6 operators have parts that contain three fields and three derivatives, e.g., $\displaystyle{f^{abc} F_\mu^{a\nu}F_\nu^{b\rho}F_\rho^{c\mu}}$ in $Q_{G^3}^{(1)}$. Also, dimension 8 operators have parts containing four fields with four derivatives, e.g.,$\left(F_{\mu \nu}^{a} F^{b \mu \nu}\right)\left(F_{\rho \sigma}^{a} F^{b \rho \sigma}\right)$ in $Q_{G^4}^{(3)}$. Both kinds of terms give $s^2/\Lambda^4$ growth in the amplitude, the former via an exchange diagram and the latter via a contact diagram. Therefore as discussed above, using the `amplitude analysis' we can constrain squares of $c_6$ and $c_6'$ relative to $c_8^{(i)}$.\\
 We calculate scattering amplitude for $gg\rightarrow gg$, written explicitly as\\
 $|p_1,\epsilon_1,a;p_2,\epsilon_2,b \rangle  \rightarrow |p_3,\epsilon_3,d;p_4,\epsilon_4,e \rangle $ (where $p$'s are the momenta, $\epsilon$'s are polarizations and Latin indices denote color of particles), at tree level getting contribution from Feynman diagrams in fig.~\ref{gluon}.
To relate  the imaginary part of amplitude to the cross-section (optical theorem), one needs to take identical initial and final states; therefore, we consider $a=d$ and $b=e$ for our calculations.

Below, we just present those terms that give $s^2$ contribution to $\mathcal{A}(s)$\footnote{We took help of the FeynCalc \cite{Feyncalc1,Feyncalc2,Feyncalc3} package to verify our calculation.}
, for reasons mentioned above. 
\begin{align}\label{3.7}
   \mathcal{A}(s)&= \mathcal{M}(s,t)|_{t\rightarrow0}\\\nn
   &=\bigg[\Big\{9f^{abc}f^{ab}_{\;\;\;c}\left(\color{red}c_6'^2-c_6^2\color{black}\right)+8\delta^{ab}\l(\color{red}c_8^{(1)}-c_8^{(2)}\color{black}\r)+4(1+\delta^{ab})\color{black}\left(\color{red}c_8^{(3)}-c_8^{(4)}\color{black}\right)\\\nn
   &\hspace{5mm}+8d^{abc}d^{ab}_{\;\;\;c}\left(\color{red}c_8^{(7)}-c_8^{(8)}\color{black}\right)\Big\}\times\Big\{\color{blue}|s\epsilon_1\cdot\epsilon_2^*-2\;\epsilon_1\cdot p_2\epsilon_2^*\cdot p_1|^2 +|s\epsilon_1\cdot\epsilon_2-2\;\epsilon_1\cdot p_2\epsilon_2\cdot p_1|^2\color{black}\Big\}\\\nn
   %
   %
   &\hspace{0.2cm}-2\Big\{9f^{abc}f^{ab}_{\;\;\;c}\; \color{red}c_6'^2\color{black}-8\delta^{ab}\color{red}c_8^{(2)}\color{black}-4(1+\delta^{ab})\color{red}\;c_8^{(4)}\color{black}-8\;d^{abc}d^{ab}_{\;\;\;c}\; \color{red}c_8^{(8)}\color{black}\Big\}\times\Big\{ \color{blue}s^2 |\epsilon_1|^2|\epsilon_2|^2\color{black}\Big\}\\\nn
    &\hspace{0.2cm}+4\Big\{9f^{abc}f^{ab}_{\;\;\;c}\color{red}c_6c_6'\color{black}-4\delta^{ab}\color{red}c_8^{(5)}\color{black}-2(1+\delta^{ab})\color{red}c_8^{(6)}\color{black}-4d^{abc}d^{ab}_{\;\;\;c}\color{red}c_8^{(9)}\color{black}\Big\}\\\nn
    &\hspace{4.5cm}\times\varepsilon^{\mu\nu\rho\sigma}Re\Big\{\color{blue}(\epsilon_{1\mu}^*\epsilon_{2\nu}p_{1\rho}p_{2\sigma})(s\epsilon_1\cdot\epsilon_2^*-2\;\epsilon_1\cdot p_2\epsilon_2^*\cdot p_1)\\\nn
    &\color{blue}\hspace{6.2cm}+(\epsilon_{1\mu}^*\epsilon_{2\nu}^*p_{1\rho}p_{2\sigma})(s\epsilon_1\cdot\epsilon_2-2\;\epsilon_1\cdot p_2\epsilon_2\cdot p_1)\color{black}\Big\}\bigg]
\end{align}

It should be noted that $a$ and $b$ in the above expression are not contracted, they are colors of incoming and outgoing particles whereas $c$ (color index) and Lorentz indices are contracted.
 For the t-channel, we get $f^{adc}f^{bec}$ factor in the numerator which is zero for our particular choice of colors, $a=d$ and $b=e$. Therefore, there is no contribution from the t-channel
 and we are allowed to take the forward limit without any divergence issues due to the t-channel pole. This isn't possible for massless theories like gravity where one gets $s^2/t$ contribution from the t-channel.

\vspace{3mm}
\noindent For further analysis we work in the COM frame and we consider general complex polarizations transverse to the momentum,
\begin{align*}
    &p_1=\{E,0,0,E\} \;, \; p_2=\{E,0,0,-E\} \;\text{(COM frame)}\\
    &\epsilon_1=\epsilon_3=\{0,\alpha,\beta,0\}\;, \; \epsilon_2=\epsilon_4=\{0,\gamma,\delta,0\}
\end{align*}

\noindent Substituting momenta and polarizations in~\ref{3.7} we get,
  \begin{align}
        \mathcal{A}(s)=&4\frac{s^2}{\Lambda^4}\Big[(2\delta^{ab}\c{red} c_8^{(1)}\c{black}+(1+\delta^{ab})\c{red}c_8^{(3)}\c{black}+2d^{abc}d^{ab}_{\;\;\;c}\c{red}c_8^{(7)}\c{black})(\c{blue}|\alpha\gamma+\beta\delta|^2+|\alpha\gamma^*+\beta\delta^*|^2\c{black})\\\nn
        &\hspace{0.4cm}+(2\delta^{ab}\c{red}c_8^{(2)}\c{black}+(1+\delta^{ab})\c{red}c_8^{(4)}\c{black}+2d^{abc}d^{ab}_{\;\;\;c}\c{red}c_8^{(8)}\c{black})(\c{blue}|\alpha^*\delta-\beta^*\gamma|^2+|\alpha^*\delta^*-\beta^*\gamma^*|^2\c{black})\\\nn
    &\hspace{0.4cm}-(2\delta^{ab}\c{red}c_8^{(5)}\c{black}+(1+\delta^{ab})\c{red}c_8^{(6)}\c{black}+2d^{abc}d^{ab}_{\;\;\;c}\c{red}c_8^{(9)}\c{black})\\\nn
     &\hspace{3cm}\times Re\c{blue}\{(\alpha^*\delta-\beta^*\gamma)(\alpha\gamma^*+\beta\delta^*)+(\alpha^*\delta^*-\beta^*\gamma^*)(\alpha\gamma+\beta\delta)\}\c{black}\Big]\\\nn
    -&9\frac{s^2}{\Lambda^4}f^{abc}f^{ab}_{\;\;\;c}\Big[|\c{red}c_6\c{black}(\c{blue}\alpha\gamma+\beta\delta\c{black})-\c{red}c_6'\c{black}(\c{blue}\alpha^*\delta^*-\beta^*\gamma^*\c{black})|^2+|\c{red}c_6\c{black}(\c{blue}\alpha\gamma^*+\beta\delta^*\c{black})-\c{red}c_6'\c{black}(\c{blue}\alpha^*\delta-\beta^*\gamma\c{black})|^2\Big]
    \end{align}
where $Re(z)$ denotes the real part of z.
From the above expression we can see that contribution from $L^{(6)}$ is always negative which means that if we don't consider dimension 8 gluonic operators in our EFT then from dispersion relation, we'll get
$$-9\frac{s^2}{\Lambda^4}f^{abc}f^{ab}_{\;\;\;c}\Big[|\c{red}c_6\c{black}(\c{blue}\alpha\gamma+\beta\delta\c{black})-\c{red}c_6'\c{black}(\c{blue}\alpha^*\delta^*-\beta^*\gamma^*\c{black})|^2+|\c{red}c_6\c{black}(\c{blue}\alpha\gamma^*+\beta\delta^*\c{black})-\c{red}c_6'\c{black}(\c{blue}\alpha^*\delta-\beta^*\gamma\c{black})|^2\Big] \geq 0$$
 which could be satisfied for arbitrary polarizations only when $c_6=0=c_6'$. It means that dim\hspace{1.5pt}6 gluonic operators cannot exist without the presence of higher dimensional operators in the SMEFT. In appendix~\ref{scalar-fermion}, we present another example of a dim\hspace{1.5pt}6 operator, involving scalar and fermion, with a similar conclusion. \\
 We get different expressions for $\mathcal{A}(s)$ (having $s^2$ dependence) for different polarizations and colors leading to multiple constraints on linear combinations of coefficients e.g., for $a=d=1 ,\; b=e=2$ and polarizations specified below we get the following expressions:
 \begin{center}
\renewcommand{\arraystretch}{1.5}     
\begin{tabular}{||c|c|c||}
 \hline
    $\sqrt{2}\epsilon_1$  &$\sqrt{2}\epsilon_2$ & $\mathcal{A}(s)/(s^2/\Lambda^4)$ \\
    \hline
     $\displaystyle{\{0,1,i,0\}}$ & $\displaystyle{\{0,1,-i,0\}}$ &
     $\displaystyle{(4c_8^{(4)}-9c_6'^2)+(4c_8^{(3)}-9c_6^2)}$\\
     $\displaystyle{\{0,1,1,0\}}$ & $\displaystyle{\{0,1,-1,0\}}$ &
     $\displaystyle{2(4c_8^{(4)}-9c_6'^2)}$\\
     $\displaystyle{\{0,1,1,0\}}$ & $\displaystyle{\{0,1,1,0\}}$ &
     $\displaystyle{2(4c_8^{(3)}-9c_6^2)}$\\
     $\displaystyle{\{0,\sqrt{2},0,0\}}$ & $\displaystyle{\{0,1,1,0\}}$ &
     $\displaystyle{4(c_8^{(4)}+c_8^{(3)}-c_8^{(6)})-9(c_6-c_6')^2}$\\
\hline
\end{tabular}
\end{center}
 
We can only constrain the $s\leftrightarrow u$ symmetric $2\rightarrow2$ scattering amplitudes using the amplitude analysis. Therefore, we consider real polarizations or particular combinations of helicity amplitudes, such that the scattering amplitude is $s\leftrightarrow u$ symmetric, to constrain the Wilson coefficients.
In table~\ref{tab:con}, we list all the different constraints we get on Wilson coefficients of dim\hspace{1.5pt}6 and dim\hspace{1.5pt}8 operators from considering different colors and polarizations.
Our constraints when considered together are stronger than those presented in \cite{consistentSMEFT}. They are also consistent with the expressions obtained for Wilson coefficients in \cite{num_values} from different UV theories. The inclusion of the dimension 6 operators and the resulting new constraints are the novel results of this section.

\begin{table}[h]
    \centering
    \resizebox{15.3cm}{!}{
    \renewcommand{\arraystretch}{2.7}
    \begin{tabular}{||c|c|c|C{5.9cm}|C{5.9cm}||}
    \hline
    \multirow{2}{*}{Colors} & $\sqrt{2}\epsilon_1 = \{0,1,1,0\}$ & $\sqrt{2}\epsilon_1 = \{0,1,1,0\}$ & $\sqrt{2}\epsilon_1 = \l\{0,1,-1,0\r\}$ & $\sqrt{2}\epsilon_1 = \{0,1,1,0\}$ \\
    \cline{2-5}
    & $\sqrt{2}\epsilon_2 = \{0,1,1,0\}$ &$\sqrt{2}\epsilon_2 = \{0,1,-1,0\}$ &$\sqrt{2(A+B)}\epsilon_2 = \l\{0,\sqrt{A}+\sqrt{B},\sqrt{A}-\sqrt{B},0\r\}$
    &$\sqrt{2(A+B)}\epsilon_2 = \l\{0,\sqrt{A}+\sqrt{B},\sqrt{B}-\sqrt{A},0\r\}$\\ 
    \hline
     $a=1, b=8$&
     $\displaystyle{c_8^{(3)}+\frac{2}{3}c_8^{(7)}}>0$& $\displaystyle{c_8^{(4)}+\frac{2}{3}c_8^{(8)}}>0$&
     \multicolumn{2}{c||}{
     $\displaystyle{4\l(c_8^{(3)}+\frac{2}{3}c_8^{(7)}\r)\l({c_8^{(4)}+\frac{2}{3}c_8^{(8)}}\r)}>\l(c_8^{(6)}+\frac{2}{3}c_8^{(9)}\r)^2$}\\
     $a=1, b=2$&
     $\displaystyle{c_8^{(3)}>\frac{9}{4}c_6^2}$&
     $\displaystyle{c_8^{(4)}>\frac{9}{4}c_6'^2}$&
     \multicolumn{2}{c||}{
     $\displaystyle{4\l(c_8^{(3)}-\frac{9}{4}c_6^2\r)\l(c_8^{(4)}-\frac{9}{4}c_6'^2\r)>\l(c_8^{(6)}-\frac{9}{2}c_6c_6'\r)^2}$}\\
     $a=1, b=4$&
     $\displaystyle{c_8^{(3)}+\frac{1}{2}c_8^{(7)}>\frac{9}{16}c_6^2}$&
     $\displaystyle{c_8^{(4)}+\frac{1}{2}c_8^{(8)}>\frac{9}{16}c_6'^2}$&
     \multicolumn{2}{c||}{
     $\displaystyle{4\l(c_8^{(3)}+\frac{1}{2}c_8^{(7)}-\frac{9}{16}c_6^2\r)\l(c_8^{(4)}+\frac{1}{2}c_8^{(8)}-\frac{9}{16}c_6'^2\r)>\l(c_8^{(6)}+\frac{1}{2}c_8^{(9)}-\frac{9}{8}c_6c_6'\r)^2}$}\\
     $a=4, b=8$&
     $\displaystyle{c_8^{(3)}+\frac{3}{2}c_8^{(7)}>\frac{27}{16}c_6^2}$&
     $\displaystyle{c_8^{(4)}+\frac{3}{2}c_8^{(8)}>\frac{27}{16}c_6'^2}$&
     \multicolumn{2}{c||}{
     $\displaystyle{4\l(c_8^{(3)}+\frac{3}{2}c_8^{(7)}-\frac{27}{16}c_6^2\r)\l(c_8^{(4)}+\frac{3}{2}c_8^{(8)}-\frac{27}{16}c_6'^2\r)>\l(c_8^{(6)}+\frac{3}{2}c_8^{(9)}-\frac{27}{8}c_6c_6'\r)^2}$} \\
     $a=b=1$&
     $\displaystyle{c_8^{(1)}+c_8^{(3)}+\frac{1}{3}c_8^{(7)}}>0$& 
     $\displaystyle{c_8^{(2)}+c_8^{(4)}+\frac{1}{3}c_8^{(8)}}>0$& 
     \multicolumn{2}{c||}{
     $\displaystyle{4\l(c_8^{(1)}+c_8^{(3)}+\frac{1}{3}c_8^{(7)}\r)
     \l(c_8^{(2)}+c_8^{(4)}+\frac{1}{3}c_8^{(8)}\r)}>\l(c_8^{(5)}+c_8^{(6)}+\frac{1}{3}c_8^{(9)}\r)^2$}\\
     $a=b=4$&
     $\displaystyle{c_8^{(1)}+c_8^{(3)}+c_8^{(7)}}>0$& 
     $\displaystyle{c_8^{(2)}+c_8^{(4)}+c_8^{(8)}}>0$&
     \multicolumn{2}{c||}{
     $\displaystyle{4\l(c_8^{(1)}+c_8^{(3)}+c_8^{(7)}\r)
     \l(c_8^{(2)}+c_8^{(4)}+c_8^{(8)}\r)}>\l(c_8^{(5)}+c_8^{(6)}+c_8^{(9)}\r)^2$}\\
    \hline
   \end{tabular}
    }
    \caption{The table contains the constraints on dim\hspace{1.5pt}6 and dim\hspace{1.5pt}8 operators' Wilson coefficients obtained using the amplitude analysis. $\epsilon_1$ and $\epsilon_2$ denote the polarizations of particles 1 and 2 respectively. The color of particle 1 is denoted by `$a$' and that of particle 2 by `$b$'. $A=4( 2\delta^{ab} c_8^{(1)}+(1+\delta^{ab}) c_8^{(3)}+2d^{abc}d^{ab}_{\;\;\;c}c_8^{(7)})-9f^{abc}f^{ab}_{\;\;\;c}c_6^2$; $ B= 4( 2\delta^{ab} c_8^{(2)}+(1+\delta^{ab}) c_8^{(4)}+2d^{abc}d^{ab}_{\;\;\;c}c_8^{(8)})-9f^{abc}f^{ab}_{\;\;\;c}c_6'^2$}
    \label{tab:con}
\end{table}

For future convenience, we will refer to a particular constraint in table~\ref{tab:con} using notation $C(i, j)$ where $i$ refers to the row, and $j$ to the column no. of the table; for example, $C(2,1)$ refers to the constraint $\displaystyle{c_8^{(3)}>\frac{9}{4}c_6^2}$.

The bounds in table~\ref{tab:con} together significantly reduce the allowed parameter space of the Wilson coefficients. Interestingly, from constraint C(2,1) and C(2,2) we can directly see that for the existence of dim\hspace{1.5pt}6 operators we need some specific dim\hspace{1.5pt}8 operators to be present. For example,  for $ Q_{G^3}^{(1)}$ we need $ Q_{G^4}^{(3)}$ to exist and for $Q_{G^3}^{(2)}$ we need $ Q_{G^4}^{(4)}$. This aspect was appreciated in the case of the electroweak bosons in \cite{constraintvectorboson}. More bounds involving dim 8 operators can be obtained by relaxing the elastic forward scattering limit as shown in \cite{Convex_geometry}.

\section{Superluminality}\label{superluminality}

We saw in the previous section that assuming an EFT to have a UV-completion that is Lorentz invariant and unitary one gets positivity constraints on the Wilson coefficients using $2\rightarrow2$ forward scattering amplitudes. It is well known that one can also reproduce some of these constraints by referring only to IR physics. It turns out that arbitrary values or signs of these coefficients can lead to superluminal propagation of field fluctuations over non-trivial backgrounds \cite{causality}. Therefore, instead of working with an S-matrix, we can work with classical wave propagation to derive interesting bounds on the Wilson coefficients by demanding the EFT to be compatible with causality. A priori, it is not absolutely clear whether this
analysis would give weaker or stronger bounds compared to the `amplitude analysis'.

In this section, we first look at the classical causality/subluminality analysis more closely; we investigate how this analysis is modified for massive fields.
We then proceed to apply this analysis to the Gluonic operators considered in sec.~\ref{amp}. Let us first briefly outline how superluminality analysis can be used to put bounds on the Wilson coefficients by demanding that signal velocity cannot be superluminal.

Consider the following Goldstone Lagrangian,
\begin{equation}
    \mathcal{L}=\frac{1}{2}(\partial\pi)^{2}+\frac{c_3}{\Lambda^{4}}(\partial\pi)^{4}
\end{equation}
Naively, it may seem that the 8-dimensional operator will not contribute to the free propagation of perturbations as it is not a quadratic term. However, this is only true for trivial backgrounds; for non-trivial backgrounds with non-vanishing derivatives one can get terms quadratic in perturbations, 
\begin{equation}
    \mathcal{L}\hspace{5pt}{\supset}\frac{c_3}{\Lambda^{4}}(\partial\pi_{0})^{2}(\partial\pi)^{2}
\end{equation}

\noindent where $\pi_{0}$ is the background field. One has to be slightly careful in choosing a background as it might happen that the background itself disallows wave-like (propagating) solutions in which case it would not be possible to run this analysis. Therefore, we usually choose a background whose scale of variation is much larger than that of field fluctuations, allowing for wave-like solutions.

It is also important to point out that given a general dispersion relation, one cannot always directly demand phase or group velocity to be subluminal as it is well known in the literature that they both can be superluminal while remaining in perfect agreement with causality \cite{Fox,superluminalvel,Recami_2009}. Instead, one usually demands the wavefront velocity be luminal. However, as already mentioned in sec.\ref{intro}, one has the so-called Kramers-Kronig relation \cite{G.M.Shore,Jackson:1998nia}
\begin{equation}
  n(\infty)=n(0)-\frac{2}{\pi}\int_{0}^{\infty}\frac{d\omega}{\omega}\textit{Im}\hspace{2pt}n(\omega)
\end{equation}
where $n(\omega)$ denotes the refractive index of the medium. For dissipative mediums, one has ${\rm Im}\hspace{1pt}n(\omega)>0$ which implies that the high frequency phase velocity is larger than the low frequency one. And since the wavefront velocity - infinite frequency limit of phase velocity - dictates the speed of information transfer, superluminal phase or group velocity can be related to violation of causality. However, if the medium allows for exhibiting gain (e.g., in the case of a Laser) then one can get ${\rm Im}\hspace{1pt}n(\omega)<0$ and
in that case, the low-frequency phase velocity cannot be a statement about causality.

The situation gets even more involved if we have a mass-like term in the equation of motion. For example, take the massive Klein-Gordon equation
\begin{equation}
    \partial^{2}\phi+m^{2}\phi=0
\end{equation}
From the above $\text{eq}^\text{n}$ we get the following dispersion relation and phase velocity,
\begin{align}
    \omega^{2}=|\vec{k}|^{2}+m^{2} \qquad ; \qquad v_{p}^{2}=1+\frac{m^{2}}{|\vec{k}|^{2}}
\end{align}
In this case phase velocity is not a good object to consider since $v_{p}(\infty)$=1 (preserving causality) while being  superluminal in the low energy limit (EFT regime). The group velocity on the other hand is (sub)luminal
\begin{equation}
    v_{g}=\frac{|\vec{k}|}{\sqrt{|\vec{k}|^{2}+m^{2}}}
\end{equation}
It was shown in \cite{Fox} that it is the group velocity which is equal to the signal velocity for Klein-Gordon modes with real mass. 

\subsection{Massive goldstone boson}
Consider the following Lagrangian
\begin{equation}
    \mathcal{L}=\frac{1}{2}(\partial\pi)^{2}+\frac{c_3}{\Lambda^{4}}(\partial\pi)^{4}-\frac{1}{2}m^{2}\pi^{2}-J\pi
\end{equation}
where $J$ is what sources the background, with the following EOM
\begin{equation}
    \partial^{2}\pi\left(1+\frac{4c_3(\partial\pi)^{2}}{\Lambda^{4}}\right)+\frac{8c_3}{\Lambda^{4}}(\partial_{\nu}\pi)(\partial^{\mu}\partial^{\nu}\pi)(\partial_{\mu}\pi)+m^{2}\pi+J=0
\end{equation}
The linearised EOM for the fluctuations $\xi=\pi-\pi_{0}$ with $\partial_{\mu}\pi_{0}=C
_{\mu}$(constant) reads
\begin{align}
     \partial^{2}\xi+\frac{8c_3}{\Lambda^{4}}C^{\mu}C^{\nu}\partial_{\mu}\partial_{\nu}\xi-\frac{4c_3C^{2}}{\Lambda^{4}}m^{2}\xi+m^{2}\xi=0
\end{align}
where we have assumed that the background terms on the l.h.s are canceled by the source $J$.
Taking Fourier transform we get, 
\begin{align}\label{adamEOM}
    k_{\mu}k^{\mu}=-\frac{8c_3}{\Lambda^{4}}\left(C.k\right)^{2}-\frac{4c_3C^{2}}{\Lambda^{4}}m^{2}+m^{2}
\end{align}
If the mass term is absent then subluminality demands $c_3>0$ as obtained by \cite{causality}. As mentioned in the above discussion, for a Klein-Gordon like dispersion relation it is the group velocity that should be (sub)luminal since the group velocity is the signal velocity. It might seem unclear whether this holds true here (since the dispersion relation is more general here, and not exactly Klein-Gordon like). However, with the choice of a purely space-like background, $C_{\mu}=\left(0,0,0,C_{(3)}\right)$, the dispersion relation takes the following form,
\begin{equation}
\omega^{2}=ak^{2}+m'^{2}
\end{equation}
where, $a=1-\frac{8c_3C^{2}_{(3)}}{\Lambda^{4}}$ and $m'^{2}=m^{2}\left(1+\frac{4c_3C^{2}_{(3)}}{\Lambda^{4}}\right)$.
This form is very close to Klein-Gordon since the parameter $a$ is very close to unity; therefore, one can perform the analysis assuming that the  group velocity is signal velocity.

We now consider the propagation of perturbations along the z-axis, then the expression for group velocity from  $\text{eq}^\text{n}$(\ref{adamEOM}) reads,
\begin{align}
    v_{g}=\frac{d\omega}{dk}=\frac{k+\frac{8c_3C_{(3)}}{\Lambda^{4}}(C_{(0)}\omega-C_{(3)}k)}{\sqrt{k^{2}+m^{2}-\frac{4c_3C^{2}m^{2}}{\Lambda^{4}}-\frac{8c_3}{\Lambda^{4}}(C_{(0)}\omega-C_{(3)}k)^{2}}+\frac{8c_3C_{(0)}}{\Lambda^{4}}(C_{(0)}\omega-C_{(3)}k)}
\end{align}

We choose the background to be varying only in the z-direction i.e. $C_{\mu}=\left(0,0,0,C_{(3)}\right)$, then demanding the group velocity to be subluminal gives
\begin{equation}
    \frac{m^{2}}{k^{2}}\left(1+\frac{4c_3C_{(3)}^{2}}{\Lambda^{4}}\right)+\frac{8c_3C_{(3)}^{2}}{\Lambda^{4}}>0
\end{equation}
Taking the limit $k\gg m$, we can drop the second term in the parenthesis and get
\begin{equation}
    \frac{m^{2}}{k^{2}}+8c_3\varepsilon >0
\end{equation}
where, $0<\varepsilon=\frac{C_{(3)}^{2}}{\Lambda^{4}}\ll1$. Finally, we get
\begin{align}
    c_{3}>-\frac{m^{2}/k^{2}}{8\varepsilon}
\end{align}
This is a ratio of two small positive quantities. Therefore, one does not get a strict positivity bound in this case from the superlumianlity analysis, unlike for the massless pions. However, the amplitude analysis still gives a strict positivity bound on $c_3$ as it is unaffected by the mass term.

\subsection{Gluon field strength operators}
We now attempt to derive constraints on the Wilson coefficients of the Gluonic operators considered in sec.~\ref{amp}, using the superluminality analysis. As we will see below, here we will not need to take recourse to the Kramers-Kronig relation because we will be able to choose backgrounds where the dispersion relation takes the simplest form i.e it is non-dispersive:
\begin{align}
    \omega=v|\vec{k}|
\end{align}
 where $v$ is a constant. In this case, phase and group velocities are the same and are equal to the signal velocity.
Due to the non-abelian nature of Lagrangian, here the calculations are tedious and are given in Appendix~\ref{sluminalgluons}. Here, we just state the main results highlighting the key assumptions that went into the analysis. We'll work with one higher-dimensional operator at a time keeping the calculations and analysis easy to follow.

\vspace{1.8mm}
First, we'll take dim\hspace{1.5pt}8 operator $Q_{G^4}^{(1)} = \left(G_{\mu \nu}^{a} G^{a \mu \nu}\right)\left(G_{\rho \sigma}^{b} G^{b \rho \sigma}\right)$ in addition to the four dimensional term in the Lagrangian
$$L=-\frac{1}{4}G_{\mu\nu}^aG^{a\mu\nu}+\frac{c_8^{(1)}}{\Lambda^4}\left(G_{\mu \nu}^{a} G^{a \mu \nu}\right)\left(G_{\rho \sigma}^{b} G^{b \rho \sigma}\right)$$
giving equation of motion,
\begin{align}\label{EOM_unperturbed}
    &-\partial_\alpha G^{f,\alpha\beta}+\frac{8c_8^{(1)}}{\Lambda^4}\l(2G_{\mu\nu}^a(\partial_\alpha G^{a,\mu\nu})G^{f,\alpha\beta}+G_{\mu\nu}^a G^{a,\mu\nu}\partial_\alpha G^{f,\alpha\beta}\r)\\\nn
    =&-g_sG^{a,\beta\nu}A_\nu^hf^{afh}+\frac{8c_8^{(1)}}{\Lambda^4}g_sf^{bfj}G^a_{\mu\nu}G^{a,\mu\nu}G^{b,\beta\sigma}A_\sigma^j
\end{align}
We expand $A^{a,\mu}=A_0^{a,\mu}+h^{a,\mu}$ where we \textit{choose} background $A_0^{a,\mu}$ to be of a particular color `$a$' with $\partial^\nu A^{a,\mu}=constant$; such background also solves the equation of motion~(\ref{EOM_unperturbed}). We \textit{look} at the linearised equation of motion for the perturbation $h^{a,\mu}$ of the same color `$a$' as that of the background,
\begin{align}
     -\partial_\alpha \partial^\alpha h^{a,\beta}+\frac{32c_8^{(1)}}{\Lambda^4}F^a_{0\mu\nu}F_0^{a\alpha\beta} \partial_\alpha\partial^\mu h^{a,\nu} = 0
\end{align}
where color index `$a$' is not contracted, and we'll drop the color index for $F_0^{\mu\nu}$ terms which should be assumed to have color `$a$'. Above $\text{eq}^\text{n}$ when expanded in terms of plane waves gives,
\begin{align}
    k_\mu k^\mu=-\frac{32c_8^{(1)}}{\Lambda^4}(F_{0\mu\nu}\epsilon^\nu k^\mu)^2
\end{align}

When a similar procedure is done considering all 6 and 8 dimensional operators, we get the following equation (color index `$c$' is contracted but not `$a$'),
\begin{align}\label{DR8}
    k_\mu k^\mu=-\frac{32}{\Lambda^4}\Bigg\{&\l(c_8^{(1)}+c_8^{(3)}+d^{aa}_{\hspace{3mm}c}d^{aac}c_8^{(7)}\r)\Big(F_{0\mu\nu}\epsilon^\nu k^\mu\Big)^2\\\nn
    +&\l(c_8^{(2)}+c_8^{(4)}+d^{aa}_{\hspace{3mm}c}d^{aac}c_8^{(8)}\r)\l(\widetilde{F}_{0\mu\nu}\epsilon^\nu k^\mu\r)^2\\
    +&\l(c_8^{(5)}+c_8^{(6)}+d^{aa}_{\hspace{3mm}c}d^{aac}c_8^{(9)}\r)\l(F_{0\mu\nu}\widetilde{F}_{0\alpha\beta}\epsilon^\nu\epsilon^\beta k^\mu k^\alpha\r)\Bigg\}\nn
\end{align}

If we look at the perturbations of the same color as that of the background, then we are effectively considering only the abelian part of our theory. Since there is no analog of dim\hspace{1.5pt}6 operators $Q^{(1)}_{G^3}$ and $Q^{(2)}_{G^3}$ in the abelian gauge theory, there is no contribution from dim\hspace{1.5pt}6 operators towards the wave propagation. However, we still see signs of the existence of different gluon colors in the form of $d^{aa}_{\hspace{3mm}c}d^{aac}$ factors which give different values for different colors in consideration.

\noindent By choosing particular background and polarization for the perturbation, we can get different bounds on Wilson coefficients $c_8^{(i)}$ using dispersion relation (\ref{DR8}). Consider the z-axis along the direction of propagation of perturbation, perturbation of the same color as that of background let say 1) with polarization, $\epsilon=\{0,1,1,0\}/\sqrt{2}$. \textit{Choose} the background such that only non-zero components of $F^{\mu\nu}$ are $F^{01}=-F^{10}$ and $F^{02}=-F^{20}$ (we have dropped $0$ from $F_0^{\mu\nu}$ to avoid confusion with time component), then we get
\begin{align}\label{7.2}
    k_\mu k^\mu=-\frac{16}{\Lambda^4}\Bigg\{&\l(c_8^{(1)}+c_8^{(3)}+\frac{1}{3}c_8^{(7)}\r)\omega^2\Big(F_{01}+F_{02}\Big)^2\\\nn
    +&\l(c_8^{(2)}+c_8^{(4)}+\frac{1}{3}c_8^{(8)}\r)\l(k^{3}\r)^2\Big(F_{02}-F_{01}\Big)^2\\\nn
    +&\l(c_8^{(5)}+c_8^{(6)}+\frac{1}{3}c_8^{(9)}\r)\omega k^3\l(F_{02}+F_{01}\r)\l(F_{02}-F_{01}\r)\Bigg\}
\end{align}
Now, if we take $F_{01}=F_{02}$ then for perturbations to be causal we need 
\begin{align}
    c_8^{(1)}+c_8^{(3)}+\frac{1}{3}c_8^{(7)}\geq0
\end{align}
which is same as the constraint $C(5,1)$ obtained in sec~\ref{amp}. Similarly by choosing different polarization and background, explicitly given in the table \ref{tab:con_superluminal}, we can reproduce $C(5,2)$ and $C(5,3)$.
Also, if instead of color 1 we choose color 4 for both background and perturbations then we get C(6,1), C(6,2) and C(6,3).

Now to probe dim\hspace{1.5pt}6 operators using superluminality analysis, we need to consider the background and perturbations of different colors. This makes the analysis rather involved. We'll first consider operator $Q_{G^3}^{(1)}=f^{abc} G_\mu^{a\nu}G_\nu^{b\rho}G_\rho^{c\mu}$ for which we have,
\begin{align*}
    L=-\frac{1}{4}G_{\mu\nu}^aG^{a\mu\nu}+\frac{c_6}{\Lambda^2}f^{abc} G_\mu^{a\nu}G_\nu^{b\rho}G_\rho^{c\mu}
\end{align*}
giving equation of motion,
\begin{align}
    -\partial_\alpha G^{f,\alpha\beta}+\frac{6c_6}{\Lambda^2}f^{fbc}\partial_\alpha\l(G^{c,\rho\alpha}G^{b,\beta\rho}\r)
        =-g_sG^{a,\beta\nu}A_\nu^hf^{afh}+\frac{6c_6}{\Lambda^2}g_sf^{abc}\l(f^{afh}G^{b,\nu\rho}G_{\hspace{1pt}\rho}^{c,\beta}A_\nu^h\r)
\end{align}
We again expand $A^{a,\mu}=A_0^{a,\mu}+h^{a,\mu}$ where we \textit{choose} background $A_0^{a,\mu}$ to be of particular color `$a$' with $\partial^\nu A^{a,\mu}=constant$, and \textit{look} at the linearised equation of motion for perturbation $h^{a,\mu}$ of different color `$f$',

\begin{align}
    &-\partial_\alpha\l( H^{f,\alpha\beta}+g_sf^{faj}A_0^{a,\alpha}h^{j,\beta}+g_sf^{fia}h^{i,\alpha}A_0^{a,\beta}\r)\\\nn
    &+\frac{6c_6}{\Lambda^2}f^{fba}\partial_\alpha\Big(F_{0}^{a,\rho\alpha} (H^{b,\beta}_{\hspace{4mm}\rho}+g_sf^{baj}A_0^{a,\beta}h^j_\rho+g_sf^{bia}h^{i,\beta}A^a_{0\rho})\\\nn
    &\hspace{2cm}-F_0^{a,\beta\rho}(H_{\hspace{1pt}\rho}^{b,\alpha}+g_sf^{baj}A^a_{0\rho} h^{j,\alpha}+g_sf^{bia}h^i_\rho A_0^{a,\alpha})\Big)\\\nn
    &= -g_s\l( H^{d,\beta\nu}+g_sf^{daj}A_0^{a,\beta}h^{j,\nu}+g_sf^{dia}h^{i,\beta}A_0^{a,\nu}\r)A_{0\nu}^af^{dfa}-g_sF_0^{a,\beta\nu}h_\nu^hf^{afh}\\\nn
    &+\frac{6c_6}{\Lambda^2}g_sf^{daa}\l(f^{dfh}F_0^{a,\nu\rho}F_{0\rho}^{a,\beta}h_\nu^h\r)\\\nn
    &+\frac{6c_6}{\Lambda^2}g_sf^{dba}\l(f^{dfa}(H^{b,\nu\rho}+g_sf^{baj}A_0^{a,\nu}h^{j,\rho}+g_sf^{bia}h^{i,\nu}A_0^{a,\rho})F_{0\rho}^{a,\beta}A_{0\nu}^a\r)\\\nn
    &+\frac{6c_6}{\Lambda^2}g_sf^{dac}\l(f^{dfa}F_0^{a,\nu\rho}(H_{\hspace{1pt}\rho}^{c,\beta}+g_sf^{caj}A^a_{0\rho} h^{j,\beta}+g_sf^{cia}h^i_\rho A_0^{a,\beta})A_{0\nu}^a\r)\nn
\end{align}
In the above equation and the following equations, `$f$' and `$a$' are not contracted but are free color indices and we'll again drop `$a$' from $F_0^{\mu\nu}$ terms which should always be assumed to have color `$a$'.

For further analysis, we take WKB approximation in which the scale of variation of background($r$) is much larger than that of perturbations ($\omega^{-1}$). We also \textit{choose} the background field to be arbitrarily small, then at leading order in $r\omega$ and $A_0$ we get,
\begin{align}\label{mainEOM:2}
    &-\partial_\alpha\partial^\alpha h^{f,\beta}+2g_sf^{fba}A_0^{a,\alpha}\partial_\alpha h^{b,\beta}-g_sf^{fba}A_{0\rho}\partial^\beta h^{b,\rho}\\\nn
    &+\frac{6c_6}{\Lambda^2}f^{fba}\Big(F_{0\rho\alpha} \partial^\alpha\partial^\beta h^{b,\rho}+F_0^{\beta\rho}\partial_\alpha\partial^\alpha h_\rho^b\Big) = 0
\end{align}
To get the dispersion relation for perturbation of color `f', we try to write the differential equation just in terms of color `f'. So we assume a particular solution for perturbation  
of some different color `$b$' which satisfies the EOM, 

\begin{align}\label{dim6_sol_main}
    -\partial^\alpha h^{b,\rho}+2g_sf^{bga}A_0^{a,\alpha} h^{g,\rho}-g_sf^{bga}A_{0\nu}\delta^{\alpha\rho} h^{g,\nu}+\frac{6c_6}{\Lambda^2}f^{bga}\l(F_0^{\sigma\alpha}\partial^\rho h^g_\sigma+F_0^{\rho\sigma}\partial^\alpha h_\sigma^g\r)=0
\end{align}
The above particular solution for other colors `$b$' modifies the $\text{eq}^\text{n}$(\ref{mainEOM:2}) to,
\begin{align}\label{dim_6massive}
    -\partial^\alpha\partial_\alpha h^{f,\beta}=&2g_s^2f^{fba}f^{gba}\l(2A_0^\alpha A_{0\alpha}h^{g,\beta}-\frac{3}{2}A_{0\nu} A_0^\beta h^{g,\nu}\r)\\\nn
    &+g\frac{6c_6}{\Lambda^2}f^{fba}f^{gba}\l(5F_0^{\beta\rho}A_0^\alpha\partial_\alpha h^{g,\rho}-A_{0\nu} F_0^{\rho\beta}\partial^\rho h^{g,\nu}+5F_{0\rho\alpha}A_0^{\alpha}\partial^\beta h^{g,\rho}\r)\\\nn
    &+36\frac{c_6^2}{\Lambda^4}f^{fba}f^{gba}\l(F_0^{\beta\rho}F_0^{\sigma\alpha}\partial_\alpha\partial_\rho h^g_\sigma+F^{\beta}_{0\rho}F_0^{\rho\sigma}\partial^\alpha\partial_\alpha h^g_\sigma+2F_0^{\rho\alpha}F_0^{\sigma\alpha}\partial^\beta\partial_\rho h^g_\sigma\r)
\end{align}
Since we have assumed a particular type of solution for other colors, we want to see how that affects the perturbation of color `$f$', so we try to write the differential equation just in terms of perturbation of color `$f$' as mentioned before. But it is not possible to replace the mass-like term (the term inside parenthesis in the first line) in the above equation unless we have an explicit solution for perturbations of all color $h^{g,\beta}$. So, for now, we'll assume that we can choose a particular background $A_0$ such that the mass-like term vanishes. We'll give below an explicit example of background and polarization of the perturbation where this assumption is satisfied.

In the second and third lines of the $\text{eq}^\text{n}$(\ref{dim_6massive}), when $g \neq f$ we again substitute $h^g$ in terms of other colors using the solution assumed $\text{eq}^\text{n}$(\ref{dim6_sol_main}). But this gives terms of higher order in $A_0$ or $\mathcal{O}\l(\frac{1}{\Lambda^4}\r)$ which we can ignore w.r.t the terms where $g=f$. Therefore only terms with $g=f$ survives at $\mathcal{O}\l(\frac{1}{\Lambda^4}\r)$ and leading order in $A_0$,

\begin{align}
    -\partial^\alpha\partial_\alpha h^{f,\beta}=
    &g\frac{6c_6}{\Lambda^2}f^{fba}f^{fba}\l(5F_0^{\beta\rho}A_0^\alpha\partial_\alpha h^{f,\rho}-A_{0\nu} F_0^{\rho\beta}\partial^\rho h^{f,\nu}+5F_{0\rho\alpha}A_0^{\alpha}\partial^\beta h^{f,\rho}\r)\\\nn
    +&36\frac{c_6^2}{\Lambda^4}f^{fba}f^{fba}\l(F_0^{\beta\rho}F_0^{\sigma\alpha}\partial_\alpha\partial_\rho h^f_\sigma+F^{\beta}_{0\rho}F_0^{\rho\sigma}\partial^\alpha\partial_\alpha h^f_\sigma+2F_0^{\rho\alpha}F_0^{\sigma\alpha}\partial^\beta\partial_\rho h^f_\sigma\r)
\end{align}

We expand the above equation in terms of waves with transverse polarization and consider the spatial wave vector, $\vec{\kappa}$ to be complex in general. We then get the following dispersion relation,
\begin{align}
     \omega^2 = |\vec{k}|^2+36\frac{c_6^2}{\Lambda^4}(f^{fba})^2\l(F_0^{\beta\rho} k_\rho\epsilon_\beta\r)^2
\end{align}
where $\vec{k}$ denotes the real part of the spatial wave vector.
\noindent After considering all dim\hspace{1.5pt}6 and dim\hspace{1.5pt}8 operators we get the following dispersion relation,
\begin{align}\label{DRgluon}
    \frac{k_\mu k^\mu}{4}=
    &\frac{9}{\Lambda^4}f^{afc}f^{af}_{\;\;\;c}\Big[\c{red}c_6\c{black}(\c{blue}F_0^{\mu\nu} k_\mu\epsilon_\nu\c{black})-\c{red}c_6'\c{black}(\c{blue}\widetilde{F}_0^{\alpha\beta}k_\alpha\epsilon_\beta\c{black})\Big]^2\\\nn
        &-\frac{4}{\Lambda^4}\Big[(2\delta^{af}\c{red} c_8^{(1)}\c{black}+(1+\delta^{af})\c{red}c_8^{(3)}\c{black}+2d^{afc}d^{af}_{\;\;\;c}\c{red}c_8^{(7)}\c{black})\c{blue}\l(F_0^{\mu\nu} k_\mu\epsilon_\nu\r)^2\\\nn
        &\hspace{0.7cm}+(2\delta^{af}\c{red}c_8^{(2)}\c{black}+(1+\delta^{af})\c{red}c_8^{(4)}\c{black}+2d^{afc}d^{af}_{\;\;\;c}\c{red}c_8^{(8)}\c{black})\c{blue}\l(\widetilde{F}_0^{\alpha\beta} k_\alpha\epsilon_\beta\r)^2\\\nn
    &\hspace{0.7cm}-(2\delta^{af}\c{red}c_8^{(5)}\c{black}+(1+\delta^{af})\c{red}c_8^{(6)}\c{black}+2d^{afc}d^{af}_{\;\;\;c}\c{red}c_8^{(9)}\c{black})\c{blue}\l(F_0^{\mu\nu} k_\mu\epsilon_\nu\r)\l(\widetilde{F}_0^{\alpha\beta} k_\alpha\epsilon_\beta\r)\c{black}\Big]
\end{align}
where `$f$' denotes the color of perturbation and `$a$' of the background. Note that the above dispersion relation is only valid, assuming that mass-like term in the $\text{eq}^\text{n}$(\ref{dim_6massive}) vanish. We had a similar situation in the previous section where we had a mass-like term dependent on the background field, which refrained us from getting a constraint on the Wilson coefficients of the theory. But in this case, since it depends on contracted four-vectors, it is possible to make the mass-like term zero along with non-zero derivatives of background by choosing an appropriate $A_0^\mu$.

\begin{table}[h]
    \centering
    \resizebox{15.1cm}{!}{
    \renewcommand{\arraystretch}{2.5}
    \begin{tabular}{||c|c|c||}
    \hline
    \multirow{2}{*}{Colors} & $\sqrt{2}\epsilon = \{0,1,1,0\}$ & $\sqrt{2}\epsilon = \{0,1,1,0\}$ \\
    \cline{2-3}
    &$A_{0\mu}=E\{\sqrt{2}(x+y),x+y,-(x+y),0\}$ 
    &$A_{0\mu}=E\{\sqrt{2}t,t,-t,0\}$ \\ 
    \hline
     $a=1, f=8$&
     $\displaystyle{c_8^{(3)}+\frac{2}{3}c_8^{(7)}}>0$& $\displaystyle{c_8^{(4)}+\frac{2}{3}c_8^{(8)}}>0$\\
     $a=1, f=2$&
     $\displaystyle{c_8^{(3)}>\frac{9}{4}c_6^2}$&
     $\displaystyle{c_8^{(4)}>\frac{9}{4}c_6'^2}$\\
     $a=1, f=4$&
     $\displaystyle{c_8^{(3)}+\frac{1}{2}c_8^{(7)}>\frac{9}{16}c_6^2}$&
     $\displaystyle{c_8^{(4)}+\frac{1}{2}c_8^{(8)}>\frac{9}{16}c_6'^2}$\\
     $a=4, f=8$&
     $\displaystyle{c_8^{(3)}+\frac{3}{2}c_8^{(7)}>\frac{27}{16}c_6^2}$&
     $\displaystyle{c_8^{(4)}+\frac{3}{2}c_8^{(8)}>\frac{27}{16}c_6'^2}$\\
     $a=f=1$&
     $\displaystyle{c_8^{(1)}+c_8^{(3)}+\frac{1}{3}c_8^{(7)}}>0$& 
     $\displaystyle{c_8^{(2)}+c_8^{(4)}+\frac{1}{3}c_8^{(8)}}>0$\\
     $a=f=4$&
     $\displaystyle{c_8^{(1)}+c_8^{(3)}+c_8^{(7)}}>0$& 
     $\displaystyle{c_8^{(2)}+c_8^{(4)}+c_8^{(8)}}>0$\\
    \hline
    \hline
    \multirow{2}{*}{Colors} & $\sqrt{2}\epsilon = \l\{0,1,-1,0\r\}$ & $\sqrt{2}\epsilon = \{0,1,1,0\}$ \\
    \cline{2-3}
    &$A_{0\mu}=  E\{\sqrt{2(D+B)}t, {(\sqrt{D}+\sqrt{B})}t ,
    {(\sqrt{D}-\sqrt{B})}t,0\}$
    &$A_{0\mu}=
    E\{\sqrt{2(D+B)}t,(\sqrt{D}+\sqrt{B})t, (\sqrt{B}-\sqrt{D})t,0\}$\\ 
    \hline
     $a=1, f=8$&
     \multicolumn{2}{c||}{
     $\displaystyle{4\l(c_8^{(3)}+\frac{2}{3}c_8^{(7)}\r)\l({c_8^{(4)}+\frac{2}{3}c_8^{(8)}}\r)}>\l(c_8^{(6)}+\frac{2}{3}c_8^{(9)}\r)^2$}\\
     $a=1, f=2$&
     \multicolumn{2}{c||}{
     $\displaystyle{4\l(c_8^{(3)}-\frac{9}{4}c_6^2\r)\l(c_8^{(4)}-\frac{9}{4}c_6'^2\r)>\l(c_8^{(6)}-\frac{9}{2}c_6c_6'\r)^2}$}\\
     $a=1, f=4$&
     \multicolumn{2}{c||}{
     $\displaystyle{4\l(c_8^{(3)}+\frac{1}{2}c_8^{(7)}-\frac{9}{16}c_6^2\r)\l(c_8^{(4)}+\frac{1}{2}c_8^{(8)}-\frac{9}{16}c_6'^2\r)>\l(c_8^{(6)}+\frac{1}{2}c_8^{(9)}-\frac{9}{8}c_6c_6'\r)^2}$}\\
     $a=4, f=8$&
     \multicolumn{2}{c||}{
     $\displaystyle{4\l(c_8^{(3)}+\frac{3}{2}c_8^{(7)}-\frac{27}{16}c_6^2\r)\l(c_8^{(4)}+\frac{3}{2}c_8^{(8)}-\frac{27}{16}c_6'^2\r)>\l(c_8^{(6)}+\frac{3}{2}c_8^{(9)}-\frac{27}{8}c_6c_6'\r)^2}$} \\
     $a=f=1$&
     \multicolumn{2}{c||}{
     $\displaystyle{4\l(c_8^{(1)}+c_8^{(3)}+\frac{1}{3}c_8^{(7)}\r)
     \l(c_8^{(2)}+c_8^{(4)}+\frac{1}{3}c_8^{(8)}\r)}>\l(c_8^{(5)}+c_8^{(6)}+\frac{1}{3}c_8^{(9)}\r)^2$}\\
     $a=f=4$&
     \multicolumn{2}{c||}{
     $\displaystyle{4\l(c_8^{(1)}+c_8^{(3)}+c_8^{(7)}\r)
     \l(c_8^{(2)}+c_8^{(4)}+c_8^{(8)}\r)}>\l(c_8^{(5)}+c_8^{(6)}+c_8^{(9)}\r)^2$}\\
    \hline
   \end{tabular}
    }
    \caption{The table contains the constraints on dim\hspace{1.5pt}6 and dim\hspace{1.5pt}8 operators' Wilson coefficients obtained using the superluminality analysis. $A_{0\mu}$ and $\epsilon$ represent the background field and polarization of the perturbation, respectively. The color of the background is denoted by `$a$' and that of perturbation by `$f$'. $D=4( 2\delta^{ab} c_8^{(1)}+(1+\delta^{ab}) c_8^{(3)}+2d^{abc}d^{ab}_{\;\;\;c}c_8^{(7)})-9f^{abc}f^{ab}_{\;\;\;c}c_6^2$; $ B= 4( 2\delta^{ab} c_8^{(2)}+(1+\delta^{ab}) c_8^{(4)}+2d^{abc}d^{ab}_{\;\;\;c}c_8^{(8)})-9f^{abc}f^{ab}_{\;\;\;c}c_6'^2$}
    \label{tab:con_superluminal}
\end{table}

We now try to get different constraints on Wilson coefficients by considering particular polarization and the background. Consider the perturbation with polarization $\epsilon=\{0,1,1,0\}/\sqrt{2}$ and choose background of the form $A_{0\mu}=E\{\sqrt{2}t,t,-t,0\}$ where E is some arbitrarily small constant. Under this configuration, mass-like term in $\text{eq}^\text{n}$(\ref{dim_6massive}) vanish and we get following non-zero components of $F_{\mu\nu}$, $F_{01}=-F_{10}$ and $F_{02}=-F_{20}$ with $F_{01}=-F_{02}$, which reduces the dispersion relation (\ref{DRgluon}) to the following form,
\begin{align}
    k_\mu k^\mu=\frac{72}{\Lambda^4}f^{afc}f^{af}_{\;\;\;c}(c'_6)^2\l(k^3\r)^2\l(F_{02}\r)^2-\frac{32}{\Lambda^4}\l(c_8^{(4)}+2d^{afc}d^{af}_{\;\;\;c}c_8^{(8)}\r)\l(k^3\r)^2\l(F_{02}\r)^2
\end{align}
and for the perturbation to be causal (dictated by the phase velocity) we require,
\begin{align}
    9f^{afc}f^{af}_{\;\;\;c}(c'_6)^2-4\l(c_8^{(4)}+2d^{afc}d^{af}_{\;\;\;c}c_8^{(8)}\r) < 0
\end{align}
We can reproduce the $C(1,2), C(2,2), C(3,2)$ and $C(4,2)$ bounds of table~\ref{tab:con} using the above relation by choosing different colors for perturbations and background. The remaining bounds of table~\ref{tab:con} can also be reproduced by choosing different background and polarization configurations, details of which have been relegated to the appendix~\ref{sluminalgluons}. In the table~\ref{tab:con_superluminal}, we present all the bounds obtained on dim\hspace{1.5pt}6 and 8 gluonic operators using the superluminality analysis by considering different configurations of the background and perturbation.

\subsection{A Non-relativistic Example: Stronger bound from Superluminality}\label{baumann}
Since we got similar bounds from both superluminality and amplitude analysis, therefore, one might think that this is always the case. Here, following \cite{Baumann_2016} we present a counter-example where superluminality gives a stronger bound.  
Consider the following Lagrangian
\begin{align}
    \mathcal{L}=\frac{1}{2}(\partial\pi)^{2}-\frac{4c_3}{3\Lambda^{2}}\dot{\pi}^{3}+\frac{2c_4}{3\Lambda^{4}}\dot{\pi}^{4}
\end{align}
This Lagrangian emerges from the EFT of Inflation \cite{inflation} Lagrangian in a particular limit \cite{Baumann_2016}. \\
The $2\rightarrow2$ forward scattering amplitude at tree level is given by
\begin{align}
\mathcal{A}(s)=\left(c_4-(2c_3)^{2}\right)\frac{s^{2}}{\Lambda^{4}}
\end{align}
Performing the `amplitude analysis' gives the following bound
\begin{align}
    c_4>(2c_3)^{2}
\end{align}
As a side remark, it is important to note that, due to subtleties related to spontaneously broken Lorentz invariance, the derivation of the above bounds in \cite{Baumann_2016} may not be completely rigorous, see \cite{Creminelli_2022} for a recent discussion.\\
Now, let us check what superluminality gives for this Lagrangian. We derive linearised EOM for the fluctuations, $\xi=\pi+\alpha t$, where $\alpha$ is a small quantity.
\begin{align}
    \ddot{\xi}+\frac{8c_3}{\Lambda^{2}}\alpha\ddot{\xi}+\frac{8c_4}{\Lambda^{4}}\alpha^{2}\ddot{\xi}-\partial_i^{2}\xi=0
\end{align}
Up to $O(\alpha)$ we have
\begin{align}
     \ddot{\xi}+\frac{8c_3}{\Lambda^{2}}\alpha\ddot{\xi}-\partial_i^{2}\xi=0
\end{align}
then the phase velocity of the fluctuation is given by
\begin{align}
    v_{phase}^{2}=1-\frac{8c_3\alpha}{\Lambda^{2}}
\end{align}
Since $\alpha$ can have any sign, therefore, the only way to preserve (sub)luminality is by taking $c_3=0$.\\
Now, up to $O(\alpha^{2})$ we have
\begin{align}
    \ddot{\xi}+\frac{8c_4\alpha^{2}}{\Lambda^{2}}\ddot{\xi}-\partial_i^{2}\xi=0
\end{align}
The phase velocity is given by
\begin{align}
      v_{phase}^{2}=1-\frac{8c_4\alpha^{2}}{\Lambda^{4}}
\end{align}
(Sub)luminality demands, $c_4\geq0$. Therefore, superluminality in this particular case gives a stronger bound than the `amplitude analysis'.

\section{Summary}\label{summary} 
We have derived constraints on dim\hspace{1.5pt}6 and dim\hspace{1.5pt}8 gluon field strength operators of SMEFT using both the amplitude and superluminality analysis. This significantly reduces the parameter space of the Wilson coefficients. Interestingly, these bounds imply that dim\hspace{1.5pt}6 operators can only exist in the presence of certain dim\hspace{1.5pt}8 operators.\\
The amplitude analysis filters out terms growing as even power of s, $s^{2n}$ where $n\geq 1$ ($n=1$ in our case). It is because of this filtering property that one is not able to put any bounds on the Wilson coefficients of dim\hspace{1.5pt}6 operators comprising four fields, e.g. $c\phi^2\partial_\mu\phi\partial^\mu\phi$ as their contribution to the tree level scattering amplitude (forward limit) grow as  $s$. But for dim\hspace{1.5pt}6 operators comprising of three fields, like some terms in $Q_{G^3}^{(1)}$ and $Q_{G^3}^{(2)}$, one indeed  gets an  $s^2$ dependence at tree level due to exchange diagrams. It is this feature that allowed us to put constraints on the square of the Wilson coefficients of dim\hspace{1.5pt}6 gluon field strength operators in SMEFT. We obtained constraints on the magnitude of dim\hspace{1.5pt}6 operators' Wilson coefficients in terms of those of dim\hspace{1.5pt}8 operators. In appendix~\ref{scalar-fermion}, we have given another example of a dim\hspace{1.5pt}6 operator (containing 3 fields) whose magnitude can be constrained in terms of dim\hspace{1.5pt}8 operators.

In the context of superluminality analysis, we have mentioned the subtleties involving the relation between low-frequency phase velocity and causality. We showed, in the case of chiral Lagrangian, that it is unclear if one gets a strict positivity bound from superluminality when the pion is considered to be massive. The reason for this is that the superluminality analysis takes into account the contribution from operators of all dimensions (unlike the `amplitude analysis'). As we have argued in our work, the contribution of the dimension 4 operator (other than the kinetic term) to the dispersion relation of the perturbation makes it unclear if one can use the phase velocity to dictate the superluminality of the perturbation. However, in the case of the gluon field strength operators, we managed to get rid of the mass-like terms in the dispersion relation by choosing specific background and polarization of the perturbation. This was possible because the mass-like term entirely depended on four-vector contractions which could be made zero despite having a non-zero field. We showed that interestingly and in a non-trivial way, the superluminality analysis for gluonic operators in the SMEFT reproduces all the  bounds obtained from the `amplitude analysis'.

The above discussion might give the impression that the amplitude analysis always gives similar or stronger bounds than the superluminality analysis. However, this is not always true. In sec.~\ref{baumann}, following \cite{Baumann_2016}, we showed an example of a non-relativistic theory (motivated by the EFT of inflation \cite{inflation}) that the superluminality gives stronger constraints than the `amplitude analysis' in this case.
Thus, we conclude that it is not clear which of the two analyses will give stronger bounds for a particular theory. Hence, ideally, one should perform both analyses in order to obtain the maximum amount of constraints on an IR effective theory.

\section*{Acknowledgments}
DG acknowledges support through the Ramanujan Fellowship and MATRICS Grant of the
Department of Science and Technology, Government of India.
DG would also like to acknowledge
support from the ICTP through the Associates Programme (2020-2025). DG thank Alex Azatov, Paolo Creminelli, and Amartya Harsh Singh for fruitful discussions.

\appendix  
  \section*{Appendix}
  \section{Details of the subluminality analysis for gluons}\label{sluminalgluons}
  \begin{itemize}
      \item For operator $Q_{G^4}^{(1)} = \left(G_{\mu \nu}^{a} G^{a \mu \nu}\right)\left(G_{\rho \sigma}^{b} G^{b \rho \sigma}\right)$ we have, 
      $$L=-\frac{1}{4}G_{\mu\nu}^aG^{a\mu\nu}+\frac{c_8^{(1)}}{\Lambda^4}\left(G_{\mu \nu}^{a} G^{a \mu \nu}\right)\left(G_{\rho \sigma}^{b} G^{b \rho \sigma}\right)$$
      Applying Euler's Lagrange equation,
      $$\partial_\alpha\frac{\partial L}{\partial\l(\partial_\alpha A^f_\beta\r)}-\frac{\partial L}{\partial A_\beta^f}=0$$ 
      for the above Lagrangian, we get EOM as,
    \begin{align}
        &-\partial_\alpha G^{f,\alpha\beta}+\frac{8c_8^{(1)}}{\Lambda^4}\l(2G_{\mu\nu}^a(\partial_\alpha G^{a,\mu\nu})G^{f,\alpha\beta}+G_{\mu\nu}^a G^{a,\mu\nu}\partial_\alpha G^{f,\alpha\beta}\r)\\\nn
        =&-g_sG^{a,\beta\nu}A_\nu^hf^{afh}+\frac{8c_8^{(1)}}{\Lambda^4}g_sf^{bfj}G^a_{\mu\nu}G^{a,\mu\nu}G^{b,\beta\sigma}A_\sigma^j
    \end{align}
    We expand $A^{a,\mu}=A_0^{a,\mu}+h^{a,\mu}$ where we \textit{choose} background $A_0^{a,\mu}$ of particular color `$a$' having $\partial^\nu A^{a,\mu}=constant$, and \textit{look} at the linearised equation of motion for perturbation $h^{a,\mu}$ also of same color `$a$', then we get
    \begin{align}
        -\partial_\alpha H^{f,\alpha\beta}+\frac{8c_8^{(1)}}{\Lambda^4}\Big(&2G_{0\mu\nu}^a\big(\partial_\alpha H^{a,\mu\nu}+g_sf^{a1c}A_0^{1,\mu}h^{c,\nu}+g_sf^{ac1}A_0^{1,\nu}h^{c,\mu}\big)G_0^{f,\alpha\beta}\\\nn
        &+G_{0\mu\nu}^a G_0^{a,\mu\nu}\partial_\alpha H^{f,\alpha\beta}\Big)=0
    \end{align}
      
    where f=a, $\displaystyle{G^{a}_{0\mu\nu}=\partial_\mu A^{a}_{0\nu}-\partial_\nu A_{0\mu}^a+g_{s}  f^{abc}A^b_{0\mu} A_{0\nu}^c}$ and $H^{a}_{\mu\nu}=\partial_\mu h^{a}_{\nu}-\partial_\nu h_{\mu}^a$ \\
    since $A^a_0$ is non zero only for particular `$a$', $\displaystyle{G^{a}_{0\mu\nu}=F^{a}_{0\mu\nu}=\partial_\mu A^{a}_{0\nu}-\partial_\nu A_{0\mu}^a}$ and is also non-zero only for color index= `$a$'. From now we'll stop writing color index for background for convenience as it is fixed to be `$a$'.
    \begin{align}
        -\partial_\alpha H^{a,\alpha\beta}+\frac{8c_8^{(1)}}{\Lambda^4}\Big(2F_{0\mu\nu}\partial_\alpha H^{a,\mu\nu}F_0^{\alpha\beta}+F_{0\mu\nu} F_0^{\mu\nu}\partial_\alpha H^{a,\alpha\beta}\Big)=0
    \end{align}
  Since we are working in Lorentz gauge, $\partial_\alpha h^\alpha=0$ , then writing $H^{a,\mu\nu}$ explicitly 
\begin{align}
     -\partial_\alpha \partial^\alpha h^{a,\beta}\l(1-\frac{8c_8^{(1)}}{\Lambda^4}F_{0\mu\nu} F_0^{\mu\nu}\r)+\frac{32c_8^{(1)}}{\Lambda^4}F_{0\mu\nu}F_0^{\alpha\beta} \partial_\alpha\partial^\mu h^{a,\nu} =0
\end{align}

Rescaling $\displaystyle{h^{a,\mu}\rightarrow\frac{ h^{a,\mu}}{\l(1-\frac{8c_8^{(1)}}{\Lambda^4}F_{0\mu\nu} F_0^{\mu\nu}\r)}}$ and considering terms only up to $\displaystyle{\mathcal{O}\l(\frac{1}{\Lambda^4}\r)}$,
\begin{align}
     -\partial_\alpha \partial^\alpha h^{a,\beta}+\frac{32c_8^{(1)}}{\Lambda^4}F_{0\mu\nu}F_0^{\alpha\beta} \partial_\alpha\partial^\mu h^{a,\nu} = 0
\end{align}
    
Taking the Fourier transform and multiplying the $\text{eq}^\text{n}$ by normalized polarization of perturbation: $\epsilon_\beta$,
\begin{align}
    k^2\cdot\epsilon_\beta \tilde{h}^{a,\beta} =\frac{32c_8^{(1)}}{\Lambda^4}F_{0\mu\nu}F_0^{\alpha\beta}\epsilon_\beta k_\alpha k^\mu \tilde{h}^{a,\nu}
\end{align}

Also, we can write $\tilde{h}^{a,\nu}=-\epsilon^{\nu}\hspace{2pt}\tilde{h}^{a,\rho}\epsilon_\rho$ considering polarization to be transverse and therefore having components only in spatial direction ($\Rightarrow\epsilon_\nu\epsilon^\nu=-1$)
\begin{align}
    k^2=-\frac{32c_8^{(1)}}{\Lambda^4}(F_{0\mu\nu}\epsilon^\nu k^\mu)^2
\end{align}

Doing the above calculation for other operators gives $\text{eq}^\text{n}$(\ref{DR8}).
\item For operator $Q_{G^3}^{(1)}=f^{abc} G_\mu^{a\nu}G_\nu^{b\rho}G_\rho^{c\mu}$,
\begin{align*}
    L=-\frac{1}{4}G_{\mu\nu}^aG^{a\mu\nu}+\frac{c_6}{\Lambda^2}f^{abc} G_\mu^{a\nu}G_\nu^{b\rho}G_\rho^{c\mu}
\end{align*}
we get EOM as,
\begin{align}
    -\partial_\alpha G^{f,\alpha\beta}+\frac{6c_6}{\Lambda^2}f^{fbc}\partial_\alpha\l(G^{c,\rho\alpha}G^{b,\beta}_{\hspace{4mm}\rho}\r)
        =-g_sG^{a,\beta\nu}A_\nu^hf^{afh}+\frac{6c_6}{\Lambda^2}g_sf^{abc}\l(f^{afh}G^{b,\nu\rho}G_{\hspace{1pt}\rho}^{c,\beta}A_\nu^h\r)
\end{align}
We again expand $A^{a,\mu}=A_0^{a,\mu}+h^{a,\mu}$ where we \textit{choose} background $A_0^{a,\mu}$ of particular color `$a$' with $\partial^\nu A^{a,\mu}=constant$, and \textit{look} at the linearised equation of motion for perturbation $h^{a,\mu}$ of color f.\\
Then EOM takes the following form for color f,
\begin{align}
    &-\partial_\alpha\l( H^{f,\alpha\beta}+g_sf^{faj}A_0^{a,\alpha}h^{j,\beta}+g_sf^{fia}h^{i,\alpha}A_0^{a,\beta}\r)\\\nn
    &+\frac{6c_6}{\Lambda^2}f^{fba}\partial_\alpha\Big(F_{0}^{a,\rho\alpha} (H^{b,\beta}_{\hspace{4mm}\rho}+g_sf^{baj}A_0^{a,\beta}h^j_\rho+g_sf^{bia}h^{i,\beta}A^a_{0\rho})\\\nn
    &\hspace{2cm}-F_0^{a,\beta\rho}(H_{\hspace{1pt}\rho}^{b,\alpha}+g_sf^{baj}A^a_{0\rho} h^{j,\alpha}+g_sf^{bia}h^i_\rho A_0^{a,\alpha})\Big)\\\nn
    &= -g_s\l( H^{d,\beta\nu}+g_sf^{daj}A_0^{a,\beta}h^{j,\nu}+g_sf^{dia}h^{i,\beta}A_0^{a,\nu}\r)A_{0\nu}^af^{dfa}-g_sF_0^{a,\beta\nu}h_\nu^hf^{afh}\\\nn
    &+\frac{6c_6}{\Lambda^2}g_sf^{daa}\l(f^{dfh}F_0^{a,\nu\rho}F_{0\rho}^{a,\beta}h_\nu^h\r)\\\nn
    &+\frac{6c_6}{\Lambda^2}g_sf^{dba}\l(f^{dfa}(H^{b,\nu\rho}+g_sf^{baj}A_0^{a,\nu}h^{j,\rho}+g_sf^{bia}h^{i,\nu}A_0^{a,\rho})F_{0\rho}^{a,\beta}A_{0\nu}^a\r)\\\nn
    &+\frac{6c_6}{\Lambda^2}g_sf^{dac}\l(f^{dfa}F_0^{a,\nu\rho}(H_{\hspace{1pt}\rho}^{c,\beta}+g_sf^{caj}A^a_{0\rho} h^{j,\beta}+g_sf^{cia}h^i_\rho A_0^{a,\beta})A_{0\nu}^a\r)\nn
\end{align}

For this, we work in WKB approximation where the scale of variation of background(r) is much larger than that of perturbations ($\omega^{-1}$), then at order $g_s$, $g_s^2$, 1/$\Lambda^2$ and leading order of r$\omega$ in each of them,
\begin{align}
    &-\partial_\alpha\l( H^{f,\alpha\beta}+g_sf^{faj}A_0^{a,\alpha}h^{j,\beta}+g_sf^{fia}h^{i,\alpha}A_0^{a,\beta}\r)\\\nn
    &+\frac{6c_6}{\Lambda^2}f^{fba}\partial_\alpha\Big(F_{0}^{a,\rho\alpha} (H^{b,\beta}_{\hspace{4mm}\rho}+g_sf^{baj}A_0^{a,\beta}h^j_\rho+g_sf^{bia}h^{i,\beta}A^a_{0\rho})\\\nn
    &\hspace{2cm}-F_0^{a,\beta\rho}(H_{\hspace{1pt}\rho}^{b,\alpha}+g_sf^{baj}A^a_{0\rho} h^{j,\alpha}+g_sf^{bia}h^i_\rho A_0^{a,\alpha})\Big)\\\nn
    &= -g_s\l( H^{d,\beta\nu}+g_sf^{daj}A_0^{a,\beta}h^{j,\nu}+g_sf^{dia}h^{i,\beta}A_0^{a,\nu}\r)A_{0\nu}^af^{dfa}\\\nn
    &+\frac{6c_6}{\Lambda^2}g_sf^{dba}f^{dfa}H^{b,\nu\rho}F_{0\rho}^{a,\beta}A_{0\nu}^a
    +\frac{6c_6}{\Lambda^2}g_sf^{dac}f^{dfa}F_0^{a,\nu\rho}H_{\hspace{1pt}\rho}^{c,\beta}A_{0\nu}^a\\\nn
\end{align}
 In above $\text{eq}^\text{n}$ since we don't know the relative order of $g_s$ and 1/$\Lambda^2$, we ignore only those terms which are definitely of less order than $g_s$, $g_s^2$, 1/$\Lambda^2$  like $g_s^2/\Lambda^2$. Also, in the above equation and all further equations `$f$' and `$a$' are not contracted but are considered particular color indices and we'll also drop `$a$' from background terms.

Lorentz gauge, $\partial_\alpha h^\alpha = 0$ implies
\begin{align}
    &-\partial_\alpha\partial_\alpha h^{f,\beta}-2g_sf^{faj}A_0^{a,\alpha}\partial_\alpha h^{j,\beta}+g_sf^{faj}A_{0\nu}\partial^\beta h^{j,\nu}\\\nn
    &+g_s^2f^{dfa}f^{daj}\l(A_0^\beta h^{j,\nu}-h^{j,\beta}A_0^\nu\r)A_0^\nu\\\nn
    &+\frac{6c_6}{\Lambda^2}f^{fba}\Big(F_{0\rho\alpha} \partial^\alpha\partial^\beta h^{b,\rho}+F_0^{\beta\rho}\partial_\alpha\partial^\alpha h_\rho^b\Big)\\\nn
    &+\frac{6c_6}{\Lambda^2}g_sf^{fba}\Big(f^{baj}A_0^\beta F_0^{\rho\alpha}\partial_\alpha h_\rho^j-f^{baj}F_0^{\rho\alpha}A_{0\rho}\partial_\alpha h^{j,\beta}\\\nn
    &\hspace{2.5cm}+f^{baj}A_0^{\alpha}F_0^{\beta\rho}\partial_\alpha h_\rho^f+f^{baj}H^{j,\nu\rho}F_{0\hspace{2pt}\rho}^{\beta}A_{0\nu}+f^{bac}F_0^{\nu\rho}H_\rho^{c,\hspace{1pt}\beta}A_{0\nu}\Big)=0
\end{align}

We can also choose the amplitude (max$|A_0(x^\mu)|$) of the background to be arbitrary small without affecting other quantities, which would make the terms of order $A_0^2$ less relevant in comparison to terms of lower order,
\begin{align}\label{EOM:2}
    &-\partial_\alpha\partial_\alpha h^{f,\beta}+2g_sf^{fba}A_0^{a,\alpha}\partial_\alpha h^{b,\beta}-g_sf^{fba}A_{0\rho}\partial^\beta h^{b,\rho}\\\nn
    &+\frac{6c_6}{\Lambda^2}f^{fba}\Big(F_{0\rho\alpha} \partial^\alpha\partial^\beta h^{b,\rho}+F_0^{\beta\rho}\partial_\alpha\partial^\alpha h_\rho^b\Big) = 0
\end{align}

Now for also some other color `$b$' we'll have similar wave equation,
\begin{align}\label{EOM:3}
    &-\partial^\alpha\partial_\alpha h^{b,\rho}+2g_sf^{bga}A_0^{a,\alpha}\partial_\alpha h^{g,\rho}-g_sf^{bga}A_{0\sigma}\partial^\rho h^{g,\sigma}\\\nn
    &+\frac{6c_6}{\Lambda^2}f^{bga}\l(F_0^{\sigma\alpha}\partial_\alpha\partial^\rho h^g_\sigma+F_0^{\rho\sigma}\partial^\alpha\partial_\alpha h_\sigma^g\r)=0
\end{align}

We consider the following solution of (\ref{EOM:3}),
\begin{align}\label{dim6_sol}
    -\partial^\alpha h^{b,\rho}+2g_sf^{bga}A_0^{a,\alpha} h^{g,\rho}-g_sf^{bga}A_{0\nu}\delta^{\alpha\rho} h^{g,\nu}+\frac{6c_6}{\Lambda^2}f^{bga}\l(F_0^{\sigma\alpha}\partial^\rho h^g_\sigma+F_0^{\rho\sigma}\partial^\alpha h_\sigma^g\r)=0
\end{align}
 and substitute in (\ref{EOM:2}),
  \begin{align}\label{dim6:massive_app}
    -\partial^\alpha\partial_\alpha h^{f,\beta}=&2g_s^2f^{fba}f^{gba}\l(2A_0^\alpha A_{0\alpha}h^{g,\beta}-\frac{3}{2}A_{0\nu} A_0^\beta h^{g,\nu}\r)\\\nn
    &+g\frac{6c_6}{\Lambda^2}f^{fba}f^{gba}\l(5F_0^{\beta\rho}A_0^\alpha\partial_\alpha h^{g,\rho}-A_{0\nu} F_0^{\rho\beta}\partial^\rho h^{g,\nu}+5F_{0\rho\alpha}A_0^{\alpha}\partial^\beta h^{g,\rho}\r)\\\nn
    &+36\frac{c_6^2}{\Lambda^4}f^{fba}f^{gba}\l(F_0^{\beta\rho}F_0^{\sigma\alpha}\partial_\alpha\partial_\rho h^g_\sigma+F^{\beta}_{0\rho}F_0^{\rho\sigma}\partial^\alpha\partial_\alpha h^g_\sigma+2F_0^{\rho\alpha}F_0^{\sigma\alpha}\partial^\beta\partial_\rho h^g_\sigma\r)
\end{align}
Now we try to get the differential equation just in terms of perturbation of color `$f$' assuming that we can choose a particular background $A_0$ such that mass term vanishes. Then in the above equation in second and third terms of r.h.s, we write $h^g$ in terms of other colors. When $g\neq f$ we get terms of higher order in $A_0$ or $\mathcal{O}\l(\frac{1}{\Lambda^4}\r)$ from (\ref{EOM:3}), therefore only $g=f$ survives at $\mathcal{O}\l(\frac{1}{\Lambda^4}\r)$ and leading order in $A_0$.
\begin{align*}
    -\partial^\alpha\partial_\alpha h^{f,\beta}=
    &g\frac{6c_6}{\Lambda^2}f^{fba}f^{fba}\l(5F_0^{\beta\rho}A_0^\alpha\partial_\alpha h^{f,\rho}-A_{0\nu} F_0^{\rho\beta}\partial^\rho h^{f,\nu}+5F_{0\rho\alpha}A_0^{\alpha}\partial^\beta h^{f,\rho}\r)\\\nn
    +&36\frac{c_6^2}{\Lambda^4}f^{fba}f^{fba}\l(F_0^{\beta\rho}F_0^{\sigma\alpha}\partial_\alpha\partial_\rho h^f_\sigma+F^{\beta}_{0\rho}F_0^{\rho\sigma}\partial^\alpha\partial_\alpha h^f_\sigma+2F_0^{\rho\alpha}F_0^{\sigma\alpha}\partial^\beta\partial_\rho h^f_\sigma\r)
\end{align*}
Taking the Fourier transform and multiplying by normalized polarization of perturbation color `$f$' $\epsilon_\beta$,
\begin{align}
    k^\mu k_\mu (\epsilon_\beta \widetilde{h}^{f,\beta})=
    &ig\frac{6c_6}{\Lambda^2}f^{fba}f^{fba}\epsilon_\beta\l(5F_0^{\beta\rho}A_0^\alpha k_\alpha \widetilde{h}^{f,\rho}-A_{0\nu} F_0^{\rho\beta}k^\rho \widetilde{h}^{f,\nu}+5F_{0\rho\alpha}A_0^{\alpha}k^\beta \widetilde{h}^{f,\rho}\r)\\\nn
    -&36\frac{c_6^2}{\Lambda^4}f^{fba}f^{fba}\epsilon_\beta\l(F_0^{\beta\rho}F_0^{\sigma\alpha}k_\alpha k_\rho \widetilde{h}^f_\sigma+F^{\beta}_{0\rho}F_0^{\rho\sigma}k^\alpha k_\alpha \widetilde{h}^f_\sigma+2F_0^{\rho\alpha}F_0^{\sigma\alpha}k^\beta k_\rho \widetilde{h}^f_\sigma\r)
\end{align}
We consider polarization to be transverse and since we can write $\tilde{h}^{f,\nu}=-\epsilon^{\nu}\hspace{2pt}\tilde{h}^{f,\rho}\epsilon_\rho$, we get
\begin{align}
     k^\mu k_\mu\l(1-36\frac{c_6^2}{\Lambda^4}(f^{fba})^2F^{\beta}_{0\rho}F_0^{\rho\sigma}\epsilon_\beta\epsilon_\sigma\r)=
    -&ig\frac{6c_6}{\Lambda^2}(f^{fba})^2\epsilon_\beta\l(5F_0^{\beta\rho}A_0^\alpha k_\alpha \epsilon^\rho-A_{0\nu} F_0^{\rho\beta}k^\rho \epsilon^\nu\r)\\\nn
    +&36\frac{c_6^2}{\Lambda^4}(f^{fba})^2\l(F_0^{\beta\rho}F_0^{\sigma\alpha}k_\alpha k_\rho\epsilon_\beta \epsilon_\sigma\r)
\end{align}
Considering $k^\mu$ to be complex in general then for the real part of dispersion relation at leading order we get,
\begin{align}
     k^\mu k_\mu = 36\frac{c_6^2}{\Lambda^4}(f^{fba})^2\l(F_0^{\beta\rho} k_\rho\epsilon_\beta\r)^2
\end{align}
\noindent After considering all dim\hspace{1.5pt}6 and dim\hspace{1.5pt}8 operators we'll get following dispersion relation,
\begin{align}\label{DRgluon_app}
    \frac{k_\mu k^\mu}{4}=
    &\frac{9}{\Lambda^4}f^{afc}f^{af}_{\;\;\;c}\Big[\c{red}c_6\c{black}(\c{blue}F_0^{\mu\nu} k_\mu\epsilon_\nu\c{black})-\c{red}c_6'\c{black}(\c{blue}\widetilde{F}_0^{\alpha\beta}k_\alpha\epsilon_\beta\c{black})\Big]^2\\\nn
        &-\frac{4}{\Lambda^4}\Big[(2\delta^{af}\c{red} c_8^{(1)}\c{black}+(1+\delta^{af})\c{red}c_8^{(3)}\c{black}+2d^{afc}d^{af}_{\;\;\;c}\c{red}c_8^{(7)}\c{black})\c{blue}\l(F_0^{\mu\nu} k_\mu\epsilon_\nu\r)^2\\\nn
        &\hspace{0.7cm}+(2\delta^{af}\c{red}c_8^{(2)}\c{black}+(1+\delta^{af})\c{red}c_8^{(4)}\c{black}+2d^{afc}d^{af}_{\;\;\;c}\c{red}c_8^{(8)}\c{black})\c{blue}\l(\widetilde{F}_0^{\alpha\beta} k_\alpha\epsilon_\beta\r)^2\\\nn
    &\hspace{0.7cm}-(2\delta^{af}\c{red}c_8^{(5)}\c{black}+(1+\delta^{af})\c{red}c_8^{(6)}\c{black}+2d^{afc}d^{af}_{\;\;\;c}\c{red}c_8^{(9)}\c{black})\c{blue}\l(F_0^{\mu\nu} k_\mu\epsilon_\nu\r)\l(\widetilde{F}_0^{\alpha\beta} k_\alpha\epsilon_\beta\r)\c{black}\Big]
\end{align}
where `$f$' denotes the color of perturbation and `$a$' of the background; given the mass term in (\ref{dim6:massive_app}) vanish.\\
 Consider the perturbation with polarization $\epsilon=\{0,1,1,0\}/\sqrt{2}$ and choose background of the form $A_{0\mu}=E\{\sqrt{2}(x+y),x+y,-(x+y),0\}$ where E is some arbitrary small constant. Under this configuration, we have $A_{0\mu}A_0^{\mu}=0$ and $\epsilon^\mu A_{0\mu}=0$ i.e. mass-like term in $\text{eq}^\text{n}$(\ref{dim6:massive_app}) vanish. For the chosen background, we get the following non-zero components of $F_{\mu\nu}$, $F_{01}=-F_{10}$, $F_{02}=-F_{20}$ and $F_{12}=-F_{21}$ with $F_{01}=F_{02}=F_{12}/\sqrt{2}$, which reduces the dispersion relation (\ref{DRgluon_app}) to the following form,
 \begin{align}
    k_\mu k^\mu=\frac{72}{\Lambda^4}f^{afc}f^{af}_{\;\;\;c}(c_6)^2\l(\omega F_{01}\r)^2-\frac{32}{\Lambda^4}\l(c_8^{(3)}+2d^{afc}d^{af}_{\;\;\;c}c_8^{(7)}\r)\l(\omega F_{01}\r)^2
\end{align}
then by demanding perturbation to be causal, we get
\begin{align}
    9f^{afc}f^{af}_{\;\;\;c}(c_6)^2-4\l(c_8^{(3)}+2d^{afc}d^{af}_{\;\;\;c}c_8^{(7)}\r) < 0
\end{align}
Choosing different combinations of colors for perturbation and background leads to constraints C(1,1), C(2,1), C(3,1), and C(4,1).
Now, if we choose the background,\\
$A_{0\mu}=
    E\{\sqrt{2(D+B)}t,(\sqrt{D}+\sqrt{B})t, (\sqrt{B}-\sqrt{D})t,0\}$ where,\\
    $D=4( 2\delta^{ab} c_8^{(1)}+(1+\delta^{ab}) c_8^{(3)}+2d^{abc}d^{ab}_{\;\;\;c}c_8^{(7)})-9f^{abc}f^{ab}_{\;\;\;c}c_6^2$;\\
    $ B= 4( 2\delta^{ab} c_8^{(2)}+(1+\delta^{ab}) c_8^{(4)}+2d^{abc}d^{ab}_{\;\;\;c}c_8^{(8)})-9f^{abc}f^{ab}_{\;\;\;c}c_6'^2$\\
    with perturbation of the same polarization as before, then the subluminality condition gives the following constraint,
\begin{align}
    -2\sqrt{DB}<4( 2\delta^{ab} c_8^{(5)}+(1+\delta^{ab}) c_8^{(6)}+2d^{abc}d^{ab}_{\;\;\;c}c_8^{(9)})-18f^{abc}f^{ab}_{\;\;\;c}c_6c_6'
\end{align}

Similarly by choosing the background to be \\
$A_{0\mu}=
    E\{\sqrt{2(D+B)}t,(\sqrt{D}+\sqrt{B})t, (\sqrt{D}-\sqrt{B})t,0\}$ along with the polarization $\epsilon=\{0,1,-1,0\}/\sqrt{2}$, we get the following constraint
\begin{align}
    2\sqrt{DB}>4( 2\delta^{ab} c_8^{(5)}+(1+\delta^{ab}) c_8^{(6)}+2d^{abc}d^{ab}_{\;\;\;c}c_8^{(9)})-18f^{abc}f^{ab}_{\;\;\;c}c_6c_6'
\end{align}    
Combining the above two constraints we can reproduce C(1,3), C(2,3), C(3,3) and C(4,3).
\end{itemize}

\section{An example with Scalar and Fermion}\label{scalar-fermion}
In this appendix, we consider another example of operators of dimensions 6 and 8 which, when subjected to the `amplitude analysis', get relative bounds on their Wilson coefficients. Consider the following Lagrangian,
 $$L^{(6)}=\frac{c_6}{\Lambda^2}\phi\partial_\mu\Bar{\Psi}\partial^\mu\Psi \; ;\;\;\;L^{(8)}=i\frac{c_8}{\Lambda^4}(\partial^\mu\partial_\nu\phi\partial_\mu\Bar{\Psi}\gamma^\nu\Psi\phi-\partial^\mu\partial_\nu\phi\Bar{\Psi}\gamma^\nu\partial_\mu\Psi\phi)$$
 where $\Psi$ represents a fermionic field and $\phi$ is a real scalar field.
 The second term in $L^{(8)}$ has to be present for it to be hermitian. Note that, the operator $L^{(6)}$ (with $\phi$ identified as the Higgs doublet, and the normal derivatives replaced by appropriate covariant derivatives) can be written in terms of a linear combination of SMEFT operators of the Warsaw basis up to total derivatives using the EOM 
 (see $\text{eq}^\text{n}(6.4)$ of \cite{dim6}).  
 
 We calculate $2\rightarrow2$ scattering amplitude with two scalars and fermions of positive helicities, $\mathcal{M}(\phi f_2^+\rightarrow\phi f_4^+)$, at tree level up to $\displaystyle{\mathcal{O}\l(\frac{1}{\Lambda^4}\r)}$. Since the dimension 6 operator considered has \textbf{3} fields, we expect to get a contribution scaling like $c_6^2s^2$ in the amplitude $\mathcal{M}(s,t)$.
 \begin{figure}[H]
     \centering
     \includegraphics[scale=0.27]{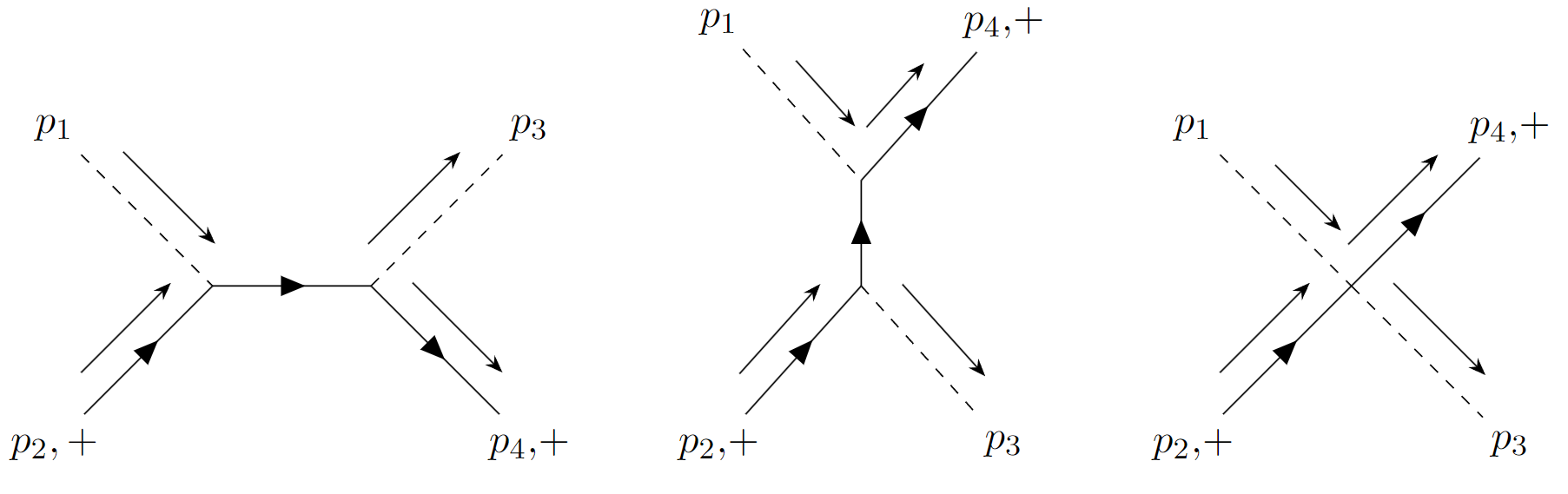}
     \caption{The first two exchange diagrams represent $s$ and $u$-channel contributions, and get contribution from $L^{(6)}$. The third contact diagram gets contribution from $L^{(8)}$.}
     \label{scal_ferm}
 \end{figure}
 \noindent The tree level amplitude gets contribution from the Feynman diagrams in figure \ref{scal_ferm} and is given by:
 \begin{align}
     \mathcal{M}(\phi f_2^+ \phi f_4^+)=& -\frac{c_6^2}{\Lambda^4}\l\{\overline{u}_{+}(p_4)(\cancel{p}_1+\cancel{p}_2)v_-(p_2)\frac{s}{4}+\overline{u}_{+}(p_4)(\cancel{p}_2-\cancel{p}_3)v_-(p_2)\frac{u}{4}\r\}\\\nn
     &+\frac{c_8}{\Lambda^4}\Big\{-\overline{u}_+(p_4)\cancel{p}_1v_-(p_2)\frac{u}{2}+\overline{u}_+(p_4)\cancel{p}_1v_-(p_2)\frac{s}{2}\\\nn
     &\hspace{1.3cm}+\overline{u}_+(p_4)\cancel{p}_3v_-(p_2)\frac{s}{2}-\overline{u}_+(p_4)\cancel{p}_3v_-(p_2)\frac{u}{2}\Big\}
 \end{align}
 which using spinor helicity formalism (for detailed introduction check \cite{scalar}) can be written as:
 $$\mathcal{M}(\phi f_2^+ \phi f_4^+)=\frac{c_8}{\Lambda^4}\Big\{\frac{-u}{2}[41]\langle12\rangle+\frac{s}{2}[41]\langle12\rangle+\frac{s}{2}[43]\langle32\rangle-\frac{u}{2}[43]\langle32\rangle\Big\}-\frac{c_6^2}{\Lambda^4}\Big\{[41]\langle12\rangle\frac{s}{4}-[43]\langle32\rangle \frac{u}{4} \Big\}$$
 
\noindent We now take the forward limit to get $\displaystyle{\mathcal{A}(s)=}$ $\displaystyle{\mathcal{M}(s,t)|_{t\rightarrow0}=\frac{s^2}{\Lambda^4}\l(2c_8-\frac{c_6^2}{2}\r)}$. We don't have to worry about t-channel pole divergence since the t-channel doesn't exist for the process considered. From positivity condition discussed in sec.~\ref{s3}, we get $$\displaystyle{4 c_8>c_6^2}$$ 
This puts an upper bound on the \textit{magnitude} of $c_6$ in terms of $c_8$ similar to what we obtained for the gluonic operators. It also implies that the 6-dimensional operator that we have considered in this example cannot exist on its own, it needs some other operator which gives a positive contribution proportional to $s^2$ in $\mathcal{A}(s)$ to survive.\\
One might not  have  expected to get $s^2$ dependence from exchange diagrams as there are only two derivatives present in $L^{(6)}$ (unlike the gluonic case which has three derivatives). However, the fermion propagator has $1/p$ dependence instead of the $1/p^2$ dependence for
gluons, and more importantly, spinors $\overline{u}$ and $v$ have implicit momentum factors. These momentum factors, in our case, manifest themselves in the form of Mandelstam variables once we take forward limit, leading to $s^2$ dependence of exchange diagrams.

\section{The arc variable}\label{arc}
\begin{figure}
    \centering
    \includegraphics[scale=0.5]{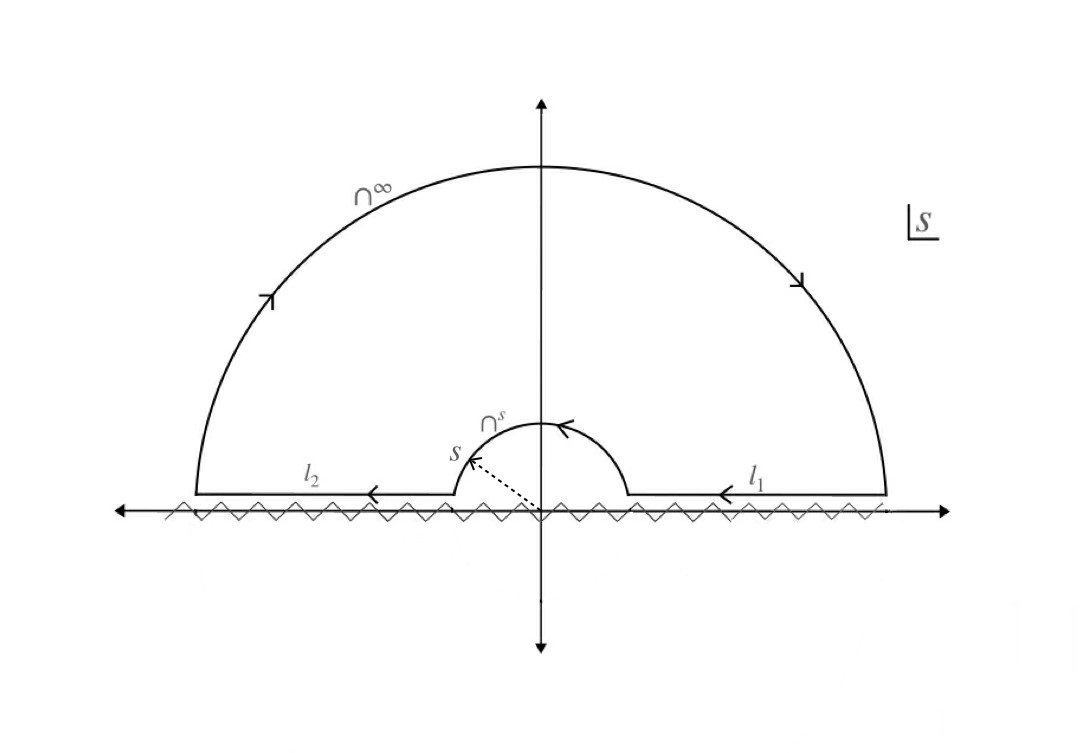}
    \caption{Contour $C$ in s-complex plane where $s$ represents some energy scale such that $\Lambda_{\text{qcd}} < s\ll\Lambda$.}
    \label{contour_arc}
\end{figure}
In sec~\ref{sec3}, we derived the constraints by calculating the 
residue at $\lim_{m^2\to0}(s\sim m^2) \to0$. However, since QCD is
confined at low energies
it would be preferable to employ a method that circumvents the need to
calculate the residue at $s\sim0$.


%


To do this, one can define the arc variable \cite{2011.00037}
\begin{equation}\label{arc_variable}
    a(s)\equiv\int_{\cap_{s}}\frac{ds'}{\pi i}\frac{\mathcal{M}(s')}{s'^{3}}
\end{equation}
where $\cap_{s}$ represents a counterclockwise semicircular path as shown in
figure \ref{contour_arc}.
Also, the Cauchy theorem implies that the integral over the contour $C=\cap_{s}+\cap_\infty+\cap_{l_1}+\cap_{l_2}$ vanishes.
Moreover, due to the Froissart bound, the integral over the arc at infinity i.e. $\cap_\infty$ vanishes. Therefore,
\begin{equation}
    a(s)=-\l[\int_{l_1} \frac{ds'}{\pi i}\frac{\mathcal{M}(s')}{s'^{3}}+\int_{l_2} \frac{ds'}{\pi i}\frac{\mathcal{M}(s')}{s'^{3}} \r]
\end{equation}
Using crossing symmetry and real analyticity, $\mathcal{M}(s+i\epsilon)=\mathcal{M}^*(-s+i\epsilon)$, we can relate the amplitude over $l_2$ to the amplitude over $l_1$, 
\begin{align}
    a(s)&=\int_{s}^{\infty}\frac{ds'}{\pi i}\frac{\mathcal{M}(s')}{s'^{3}}+\int_{-\infty}^{-s}\frac{ds'}{\pi i}\frac{\mathcal{M}(s')}{s'^{3}}\\
    &=\int_{s}^{\infty}\frac{ds'}{\pi i}\frac{\mathcal{M}(s')}{s'^{3}}-\int_{s}^{\infty}\frac{ds'}{\pi i}\frac{\mathcal{M}^*(s')}{s'^{3}}=\frac{2}{\pi}\int_{s}^{\infty}ds'\frac{\text{Im}\mathcal{M}(s')}{s'^{3}}
\end{align}
The optical theorem relates the imaginary part of amplitude to the cross-section, $\text{Im}\mathcal{M}(s')=s'\sigma(s')$,
\begin{equation}
    a(s)=\frac{2}{\pi}\int_{s}^{\infty}ds'\frac{\sigma(s')}{s'^{2}}>0
\end{equation}
We can systematically compute the arc variable, $a(s)$, as an expansion in $s$ using $\text{eq}^{n}$(\ref{arc_variable}) withing the validity of the EFT regime. For amplitude of the form, $\mathcal{M}(s)=\sum_{n=0}c_{2n}s^{2n}$, which is the case for gluon-gluon scattering, the arc variable is given by the Wilson coefficient, $a(s)=c_{2} >0$ i.e. the coefficient of $s^2$ in the amplitude is always positive.

 \nocite{*}
\color{black}
\bibliographystyle{JHEP}
{\footnotesize
\bibliography{reference}}

\end{document}